\documentclass[nonblindrev]{informs3}

\usepackage{endnotes}
\usepackage{tikz}
\usepackage{amsmath}
\usepackage{threeparttable}
\usepackage{adjustbox}
\usepackage{bbm}
\usepackage{comment}
\usepackage{xcolor}
\definecolor{mypink}{RGB}{255,0,255}
\definecolor{myred}{RGB}{255,153,153}
\definecolor{mygrey}{RGB}{128,124,127}
\definecolor{mygreen}{RGB}{55,225,57}

\usetikzlibrary{shapes}
\newcommand*\circled[1]{\tikz[baseline=(char.base)]{
        \node[shape=rounded rectangle,draw,inner sep=2pt] (char) {#1};}}
\let\footnote=\endnote
\let\enotesize=\normalsize
\def\notesname{Endnotes}%
\def\makeenmark{\hbox to1.275em{\theenmark.\enskip\hss}}
\def\enoteformat{\rightskip0pt\leftskip0pt\parindent=1.275em
  \leavevmode\llap{\makeenmark}}

\usepackage{natbib}
 \bibpunct[, ]{(}{)}{,}{a}{}{,}%
 \def\bibfont{\small}%
 \def\newblock{\ }%
 \def\BIBand{and}%

\TheoremsNumberedThrough     
\ECRepeatTheorems

\EquationsNumberedThrough    

\def\vomega{\mbox{\boldmath $\omega $}}

\begin{document}


\RUNAUTHOR{Fibich, Gillingham, and Levin}

\RUNTITLE{Boundary Effects in the Diffusion of New Products on Cartesian Networks}

\TITLE{Boundary Effects in the Diffusion of New Products on Cartesian Networks}

\ARTICLEAUTHORS{%
\AUTHOR{Gadi Fibich, Tomer Levin}
\AFF{Tel Aviv University, Tel Aviv 6997801, Israel, \EMAIL{fibich@tau.ac.il}, \EMAIL{levintmr@gmail.com}} 
\AUTHOR{Kenneth Gillingham}
\AFF{Yale University, New Haven, CT 06033, USA, \EMAIL{kenneth.gillingham@yale.edu}}
} 

\ABSTRACT{%
We analyze the effect of 
boundaries in the discrete Bass model on D-dimensional Cartesian networks. In 2D, this model describes the diffusion of new products that spread primarily by spatial peer effects, such as residential photovoltaic solar systems. We show analytically that nodes (residential units) that are located near the boundary are less likely to adopt than centrally-located ones. This boundary effect is local, and decays exponentially  with the distance from the boundary. At the aggregate level, boundary effects reduce the overall adoption level. The magnitude of this reduction scales as~$\frac{1}{M^{1/D}}$, where~$M$ is the number of nodes. Our analysis is supported by empirical evidence on the effect of boundaries on the adoption of solar.
}%



\maketitle

%


\section{Introduction}
Diffusion in networks has attracted the attention of researchers
in physics, mathematics, biology, computer science, social sciences, economics, and management science,
as it concerns the spreading of ``items'' ranging from diseases and computer viruses to rumors, information, opinions, technologies and
innovations~\citep{Albert-00,Anderson-92,Jackson-08,Pastor-Satorras-01,Rogers-03,Strang-98}. Relatedly, understanding the diffusion of new products is a central problem in marketing~\citep{Mahajan-93}.

The first mathematical model of diffusion of new products
was proposed by~\cite{Bass-69}.  In the Bass model, individuals adopt a new product because
of {\em external influences} by mass media, and {\em internal influences}
by individuals who have already adopted the product.
This seminal paper inspired a
huge body of theoretical and empirical research~\citep{Hopp-04}.
Most of this research was also carried out using compartmental models, which are typically given by deterministic  ordinary differential equations. Such models implicitly assume that
all individuals within the population are equally likely to influence each other, i.e., that the underlying social network is a complete graph.

There is considerable evidence that some products spread predominantly through a spatial peer effect, whereby geographically proximate households influence each other. In other words, if your neighbor adopts the product, you are more likely to adopt the product. These effects have been shown in contexts such as residential rooftop solar systems~\citep{barton2021decay, kraft2018credibility, rai2016overcoming, rai2015agent,wolske2018accelerating} and dry landscaping in desert climates~\citep{bollinger2020peer}.  Indeed, {\em the key predictor of a new solar installation is having a neighbor who already installed one}~\citep{Bollinger-12,Graziano-15}. As a result, the diffusion dynamics is inherently spatio-temporal, rather than just temporal as in compartmental models. A natural choice of a model for the diffusion of such products is the discrete Bass model on a two-dimensional Cartesian grid, where each node represents a
residential unit. This network is sparse, and is thus fundamentally different from the complete network structure associated with compartmental Bass models.

Discrete Bass models are stochastic particle models. As such, they are considerably harder to analyze than compartmental Bass models. The discrete Bass model on D-dimensional Cartesian grids was first analyzed by~\cite{OR-10}. That study considered infinite Cartesian networks, and finite Cartesian networks with periodic (toroidal) boundary conditions. In both cases, by translation invariance, all nodes have the same adoption probability, which leads to a huge simplification in the analysis.

Real-life residential units, however, belong to a finite network (e.g., a town), such that units (nodes) that lie at the town boundary can be influenced by fewer neighbors than centrally located units.
These 2D networks can also have {\em interior boundaries} which prevent adjacent units from influencing each other (river, highway). Interior boundaries can also be non-physical. For example, in Switzerland, there is a language boundary between French-speaking and German-speaking regions that hinders the diffusion of solar systems~\citep{carattini2018social}. The municipal boundary between adjacent towns can also influence the diffusion, if, e.g., a marketing campaign occurs in just one of the towns. The diffusion of residential rooftop solar is an especially interesting context to explore the effects of boundaries, because it is a relatively new technology that is increasingly being adopted by households in many countries, but it was only a few years ago when there were no residential rooftop solar installations. And while they are becoming more common, even today most households do not have rooftop solar,
and so {\em a clear diffusion process is underway in real-world data}.

As noted, the lack of translation invariance on finite networks makes the analysis more complex. The  {\em qualitative effect} of boundaries on the diffusion of new products was analyzed by~\cite{Bass-boundary-18}.
 In this paper, we analyze, apparently for the first time, the {\em quantitative effect} of  boundaries in the Bass model on D-dimensional Cartesian networks. Our main findings are:
\begin{enumerate}
\item The adoption probabilities of boundary nodes can be substantially lower than those of central nodes, but are at least half those of central nodes.

\item The effect of the boundary on the adoption probability of nodes is a local phenomenon, which decays exponentially fast with the distance from the boundary.
\item At the aggregate level, the boundary reduces the aggregate adoption level, relative to that on infinite networks. The magnitude of this reduction is inversely proportional to~$M^{1/D}$, where~$M$ is the network size.

\item We provide the empirical evidence for the effect of boundaries on the diffusion of residential rooftop solar. Indeed, based on our empirical data, external boundaries (municipality borders) reduce the adoption probability of near-boundary units by 30\%.

\end{enumerate}


  The paper is organized as follows. Section~\ref{sec:DBM} reviews the discrete Bass model on~$D$-dimensional Cartesian grids. The 2D case is the dimension of most relevance to applications;  the 1D case is the easiest to analyze,
   yet it captures all the essential aspects of boundary effects.
Sections~\ref{sec:local_effect_1D} and~\ref{sec:higher_dimensions_results} analyze the local effect of boundaries on the adoption probability of nodes in one-dimensional
 and in multi-dimensional networks, respectively.
Section~\ref{sec:global_effect_high_dim} concerns the global effect of boundaries
on the aggregate level of adoption.
%
Section~\ref{sec:empirical} provides empirical evidence for boundary effects in the diffusion of solar. Section~\ref{sec:discussion} discusses the robustness of the
model and provides final remarks and managerial implications.

From a  theoretical perspective, this manuscript provides the first analysis of the
quantitative effects of boundaries on the diffusion of new products. From a  managerial perspective,
our results indicate that at the municipal level, the effect of boundaries is small, and probably negligible.  At the local level, however, boundaries have a
substantial influence of lowering the adoption of the near-boundary residential units.

\section{Discrete Bass model}
\label{sec:DBM}


In this paper, we analyze the diffusion of new products in the discrete Bass model on a
$D$-dimensional Cartesian box $B_D:=[1, \dots,  M_1]^D \subset \mathbb{Z}^D$ with $M = M_1^D$ nodes.
The diffusion of rooftop solar by households corresponds to $D=2$.
Let $X_{\bf j}(t)$ denote the {\em state} of node/individual/residental-unit ${\bf j}:=\left(j_1,\ldots, j_D\right)$ at time $t$, so that
\begin{equation*}
	X_{\bf j}(t)=\begin{cases}
		1, \qquad {\rm if}\ {\bf j} \ {\rm is\ an\ adopter \ at\ time}\ t,\\
		0, \qquad {\rm otherwise,}
	\end{cases}
   \qquad {\bf j} \in B_D.
\end{equation*}
The diffusion starts when the new product is introduced into the market at time~$t=0$. Therefore, {\em initially all nodes are nonadopters}, i.e.,
\begin{subequations}
	\label{eqs:Bass-general_D}
	\begin{equation}
		\label{eq:Bass-general_D-initial}
		X_{\bf j}(0)=0, \qquad {\bf j} \in B_D.
	\end{equation}
The adoption by each node is stochastic. Thus, each nonadopter~${\bf j}$ experiences \emph{internal influences} (word of mouth, peer effects) to adopt
at the rate of~$\frac{q}{2D}$ by its neighbors who already adopted.
Let  us denote the number of adopters connected to~${\bf j}$ by
$$
N_{\bf j}(t):= \sum_{i=1}^D \left(X_{{\bf j}-\hat{\bf e}_i}(t) + X_{{\bf j}+\hat{\bf e}_i} (t)\right),
$$
where $\hat{\bf e}_i$ is the unit vector in the direction of the $i$th~coordinate, and $X_{\bf k}(t) := 0$ if ${\bf k} \not\in  B_D$.
Then the rate of
internal influences on~${\bf j}$ is~$\frac{q}{2D} N_{\bf j}(t)$.
In addition, a nonadopter~${\bf j}$ experiences \emph{external influences}  by mass media or commercials to adopt,
at the constant rate of~$p$. This assumption sets the Bass model apart from epidemiological models  (SI, SIR, \dots) on networks~\citep{kiss2017mathematics}, and has profound implications for the model robustness
with respect to long-range connections (Section~\ref{sec:discussion}).
Finally, we assume that once ${\bf j}$ adopts the product, it remains an adopter at all later times. Thus, for example, we are treating rooftop solar adoption as a one-time event, as for the most part it is (nearly all of panels on a rooftop solar system last for upwards of 25 years).
Therefore, so long that ${\bf j}$ did not adopt, his/her adoption time is exponentially distributed with rate $\lambda_{\bf j}(t) =  p+\frac{q}{2D}N_{\bf j}(t)$.
Hence, the stochastic adoption
of~${\bf j} \in B_D$  in the time interval~$(t, t+\Delta t)$
as  $\Delta t \to 0$ evolves accroding to
\begin{equation}
	\label{eq:Bass-general_D-dynamics}
	\mathbb{P} (X_{\bf j}(t+\Delta t )=1  \mid  {\bf X}(t))=
	\begin{cases}
		\left(p+\frac{q}{2D}N_{\bf j}(t) \right) \Delta t , & {\rm if}\ X_{\bf j}(t)=0,
		\\
		\qquad 1,\hfill & {\rm if}\ X_{\bf j}(t)=1,
	\end{cases}
	  \qquad {\bf j} \in B_D,
\end{equation}
\end{subequations}
where ${\bf X}(t):= \{X_{\rm j} (t)  \mid {\bf j} \in B_D \}$ are the states of the population at time $t$.

The influence rate of each adopter is defined as~$\frac{q}{2D}$ (and not, e.g., as~$q$),
so that regardless of the dimension, {\em $q$ is the maximal internal influence rate experienced by an interior node}, which is when all its $2D$ peers are adopters. This definition of~$q$ does not effect the diffusion dynamics in each dimension, but it provides for a more meaningful comparison of networks with different dimensions~\citep{OR-10}.
%


The maximal internal influence that can be experienced by  boundary nodes
is strictly below~$q$, see~\eqref{eq:Bass-general_D-dynamics},
since $N_{\bf j}(t)< 2D$.
 Therefore, the boundaries have a \emph{local} effect of lowering  the adoption probabilities
$$f_{\bf j}^{B_D}(t):=\mathbb{P}(X_{\bf j}(t)=1)$$ of the near-boundary nodes, relative to those of more central nodes. As a result, the boundaries have a \emph{global} effect of reducing the expected adoption level on~$B_D$, defined by
\begin{equation}
	\label{eq:f^B_D-def}
f^{B_D}(t):=\frac{1}{M}  \sum_{{\bf j} \in B_D} f_{\bf j}^{B_D}(t).
\end{equation}
 In this paper we study both the local and the global effects of  boundaries on the diffusion.

%

\subsection{One-dimensional networks (lines and circles)}
\label{sec:circle}

We will first analyze the one-dimensional case, which is considerably easier than the multi-dimensional case.  
{\bf The discrete Bass model on the line} $B_{\rm 1D}:=[1,\ldots,M]$ reads,
see~\eqref{eqs:Bass-general_D},
\begin{subequations}
	\label{eq:two_sided_line_all-isotropic}
	\begin{equation}
		\label{eq:initial_two_sided_line-isotropic}
		X_j(0)=0, \qquad j\in \{1, \dots, M\},
	\end{equation}
	and  as $\Delta t\to0$,
	\begin{equation}
		\label{eq:1D_Both_line-isotropic}
		\mathbb{P} (X_j(t+\Delta t)=1 \, \vert \,  {\bf X}(t))=
		\begin{cases}
			1,&  \quad {\rm if}\ X_j(t)=1,\\
			\left(p+\frac{q}{2} N_j(t) \right) \Delta t, & \quad {\rm if}\ X_j(t)=0,
		\end{cases}
		\qquad j \in\{1, \dots, M\},
	\end{equation}
where the number of adopters influencing~$j$ is
	\begin{equation}
	\label{eq:1D_N(t)}
	N_j(t) :=
	\begin{cases}
	X_2(t),&  \quad {\rm if}\ j=1,\\
		X_{j-1}(t)+ X_{j+1}(t), & \quad {\rm if}\ j=2, \dots, M-1, \\
	X_{M-1}(t),&  \quad {\rm if}\ j=M.
	\end{cases}
\end{equation}
\end{subequations}
We denote by~$f^{[1, \dots, M]}_j(t):=\mathbb{P}(X_j(t) = 1)$
the adoption probability of node~$j$,
where~$X_j(t)$ is the solution of~\eqref{eq:two_sided_line_all-isotropic},
and by $f^{[1, \dots, M]} := \frac1M \sum_{j=1}^M f^{[1, \dots, M]}_j$
 the expected adoption level on $[1, \dots, M]$.

To isolate the boundary effect in 1D, we will compare the expected adoption level on
the line with that on a circle.
{\bf The discrete Bass model on a circle}  with~$M$ nodes is also given by~\eqref{eq:two_sided_line_all-isotropic}, except that  in~\eqref{eq:1D_N(t)}, $N_1(t) = X_M(t)+X_2(t)$ and $N_M(t) = X_{M-1}(t)+X_1(t)$. This problem is easier to analyze,
since by translation invariance, the probability of node~$j$ on the circle to adopt is
independent of~$j$, i.e., $f^{\rm circle}_j(t;p,q,M) \equiv f^{\rm circle}(t;p,q,M)$ for $ j \in\{1, \dots, M\}$.
Translation invarariance also holds on the {\em infinite line}, i.e.,
 $f^{\mathbb{Z}}_j \equiv f^{\mathbb{Z}}$, where  $\mathbb{Z} := (-\infty, \dots, \infty)$.

Using translation invariance, one can obtain explicit expressions
for the infinite line and circle:
\begin{lemma}[\cite{OR-10,Bass-boundary-18}]
	\label{lem:f_1D}
The expected adoption level on the infinite line and on the infinite circle are identical, and are given by
		\begin{equation}
			\label{eq:f_1D=lim}
			\lim_{M\rightarrow \infty}f^{\rm circle}(t;p,q,M) =
			f^{\mathbb{Z}}(t;p,q)
			 = f^{\rm 1D}(t;p,q),
%
\qquad 		f^{\rm 1D}(t;p,q) := 1-e^{-\left(p+q\right)t+\frac{q}{p}\left(1-e^{-pt}\right)}.
		\end{equation}
\end{lemma}

\section{Local boundary effects in 1D}
\label{sec:local_effect_1D}

\subsection{Maximal local boundary effect}
\label{sec:sub_local_1D}

   As noted, there is no translation invariance on finite or semi-infinite lines.
 To compute the maximal effect of boundaries on the adoption probability of nodes,
we compare the two extreme cases of boundary ($j=1,M$) and central ($j \approx  \frac{M}{2}$) nodes.
In addition, to avoid multiple effects from the left and right boundaries,
we begin with the semi-infinite line
$
\mathbb{Z}^+:=\left[1,\ldots, \infty\right),
$
 which has a single boundary at $j=1$.
We denote by~$f^{\mathbb{Z}^+}_j(t):=\mathbb{P}(X_j(t) = 1)$
the adoption probability of node~$j \in \mathbb{Z}^+$,
where~$X_j(t)$ is the solution of~\eqref{eq:two_sided_line_all-isotropic}
with $M  = \infty$.
Denote the adoption probabilities of boundary and  central nodes on~$\mathbb{Z}^+$ by
\begin{equation}
	\label{eq:def-f_bdry+f_central}
	f^{\mathbb{Z}^+}_{\rm bdry}:=f^{\mathbb{Z}^+}_{j=1}, \qquad   f^{\mathbb{Z}^+}_{\rm central}:=\lim_{j\rightarrow \infty}f^{\mathbb{Z}^+}_j.
\end{equation}
We can obtain explicit expressions for these two probabilities:
%
%
\begin{lemma}
\label{lem:local_effect-semi-infinite-line}
\begin{equation}
\label{eq:local_effect-semi-infinite-line}
 f^{\mathbb{Z}^+}_{\rm bdry}(t;p,q)=f^{\rm 1D}\left(t;p,\frac{q}{2}\right), \qquad   f^{\mathbb{Z}^+}_{\rm central}(t;p,q)=  f^{\rm 1D}(t;p,q),
\end{equation}
where~$f^{\rm 1D}$ is the adoption probability on the infinite circle/line, see~\eqref{eq:f_1D=lim}.
\end{lemma}
\proof{Proof.} See Appendix~\ref{app:local_effect}.\Halmos

Thus, the change in the adoption probability from a central node to a boundary node is equivalent to reducing the peer-effect parameter from~$q$ to~$\frac{q}{2}$.

Using the explicit expressions~\eqref{eq:local_effect-semi-infinite-line}, we  show that
{\em throughout the diffusion process,   the adoption probability of the boundary node is lower than,
	but at least half, the adoption probability of a central node}:
\begin{lemma}
	\label{lem:R-boundary-1D}
     Let $p,q>0$.
	Then
	\begin{equation}
		\label{eq:R-lower-and-upper-bounds}
		\frac12 f^{\mathbb{Z}^+}_{\rm central}(t)<	f^{\mathbb{Z}^+}_{\rm bdry}(t)<f^{\mathbb{Z}^+}_{\rm central}(t),
		\qquad t>0.
	\end{equation}
\end{lemma}

\proof{Proof.} See Appendix~\ref{app:R-boundary-1D}.\Halmos

\begin{figure}[ht!]
	\begin{center}
		\scalebox{0.7}{\includegraphics{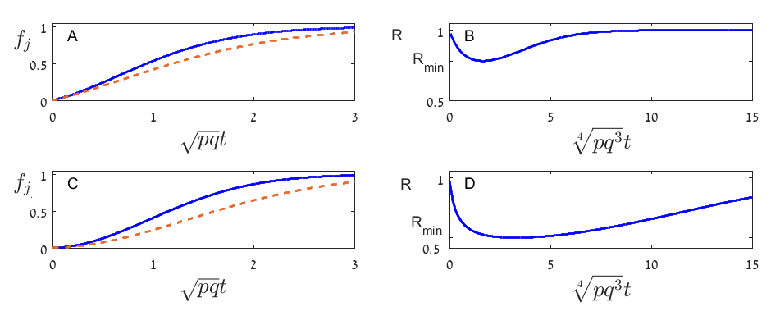}}
		\caption{(A)~The adoption probability of central ($f^{\mathbb{Z}^+}_{\rm central}$, blue solid) and  boundary ($f^{\mathbb{Z}^+}_{\rm bdry}$, red dashes) nodes on the semi-infinite line, see~\eqref{eq:local_effect-semi-infinite-line},
			as a  function of~$\sqrt{pq}t$.   Here~$\frac{q}{p} = 10$. (B)~The ratio~$R:={f^{\mathbb{Z}^+}_{\rm bdry}}/{f^{\mathbb{Z}^+}_{\rm central}}$ as a function of~$\sqrt[4]{pq^{3}}t$ for~$\frac{q}{p}=10$. (C) and (D):~Same as~(A) and~(B) for~$\frac{q}{p} = 10^3$.}
		\label{fig:1D-middle_boundary}
	\end{center}
\end{figure}

Plotting the explicit expressions~\eqref{eq:local_effect-semi-infinite-line}
in Figure~\ref{fig:1D-middle_boundary}
confirms that
$\frac12 f^{\mathbb{Z}^+}_{\rm central}<	f^{\mathbb{Z}^+}_{\rm bdry}<f^{\mathbb{Z}^+}_{\rm central}$. Intuitively, this is because central nodes can adopt due to internal influences from two infinite rays of nodes arriving from the left and from the right, whereas the boundary node can only adopt due to internal influences from a single infinite ray of nodes arriving from the right.

%

%

%

\begin{figure}[ht!]
	\begin{center}
		\scalebox{1}{\includegraphics{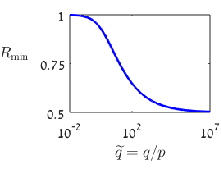}}
		\caption{$R_{\min}(\widetilde{q})$, ploted on a semi-log scale.}
		\label{fig:R_min}
	\end{center}
\end{figure}

{\em How significant are these local boundary effects?} Let
\begin{equation}
	\label{eq:R_min-def}
R_{\min}(p,q):=\min_{0 \le t<\infty} R(t;p,q), \qquad R(t;p,q):= \frac{f^{\mathbb{Z}^+}_{\rm bdry}(t;p,q)}{f^{\mathbb{Z}^+}_{\rm central}(t;p,q)} = \frac{f^{\rm 1D}\left(t;p,q\right)}{f^{\rm 1D}\left(t;p,\frac{q}{2}\right)}.
\end{equation}
Then $R=1$ when there are no boundary effects, and {\em the  maximal local boundary effect is captured by~$1-R_{\min}$}.
By Lemma~\ref{lem:R-boundary-1D},
$\frac12 \leq R_{\min}\leq 1$. In addition,
a standard dimensional argument (Appendix~\ref{app:R_min(tilde(q))}) shows that
$R_{\min} = R_{\min}(\widetilde{q})$, where $\widetilde{q}:= \frac{q}{p}$
is the ratio of internal and external influences.
Since there are no boundary effects when $q=0$, $R_{\min}(\widetilde{q}=0)=1$.
As $\widetilde{q}$ increases, internal influences become more dominant, hence so do
boundary effects, and so
$R_{\min}(\widetilde{q})$ {\em decreases monotonically}, as confirmed numerically in Figure~\ref{fig:R_min} (and compare also  Figures~\ref{fig:1D-middle_boundary}B and~\ref{fig:1D-middle_boundary}D).
In particular, the theoretical lower bound $R_{\min} =  \frac12$
is reached as $\widetilde{q}\rightarrow \infty$:
\begin{lemma}
	\label{lem:R_min=1/2}
	$\lim_{\widetilde{q}\rightarrow \infty}R_{\min}(\widetilde{q}) = \frac12$.
\end{lemma}
\proof{Proof.} See Appendix~\ref{app:R_min=1/2}.\Halmos

\subsection{Spatio-temporal decay of boundary effects}

The adoption probabilities~$\{f_j^{\mathbb{Z}^+}\}$ of nodes on the semi-infinite line are monotonically increasing with~$j$,
from~$f^{\mathbb{Z}^+}_{\rm bdry}:=f^{\mathbb{Z}^+}_{j=1}$ to~$f^{\mathbb{Z}^+}_{\rm central}:= \lim_{j \to \infty} f^{\mathbb{Z}^+}_{j} f^{\rm 1D}$:
\begin{lemma}
	\label{lem:f_j-monotone-two-sided-semi-infinite}
	$f_j^{\mathbb{Z}^+}(t;p,q)$  is monotonically  increasing in~$j$.
\end{lemma}
\proof{Proof.} See Appendix~\ref{app:f_j-monotone-two-sided-semi-infinite}.\Halmos

Hence, boundary effects decay monotonically with the distance~$j$
from the boundary. At any positive time, this decay occurs at the
 super-exponential rate of~$\frac{1}{j^j}$ as $ j \to \infty$:
\begin{theorem}
\label{thm:1D_decay}
 Let $p,q>0$.
Then for any $t>0$,
\begin{equation}
	\label{eq:1D_decay}
	0<
	f^{\rm 1D}(t;p,q) -
	f^{\mathbb{Z}^+}_{j}(t;p,q)<e^{-\left(p+\frac{q}{2}\right)t}\left(\frac{e\frac{q}{2}t}{j}\right)^j, \qquad  j \ge  \frac{q}{2}t.
\end{equation}
\end{theorem}
\proof{Proof.} See Appendix~\ref{app:1D_decay_1D}. \halmos

\begin{figure}[ht!]
	\begin{center}
		\scalebox{1.4}{\includegraphics{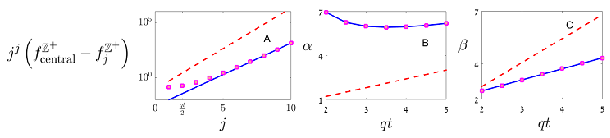}}
		\caption{(A)~$j^j \left(f^{\rm 1D}-f^{\mathbb{Z}^+}_{j}\right)$ as a function of~$j$ (\protect \tikz \protect\draw[fill=mypink] (0,0) -- (0,0.2)--(0.2,0.2)--(0.2,0)--(0,0);); $y$-axis is in log scale. Solid line is the fitted curve~$e^{-\alpha}\beta^j$ with~$\alpha \approx 6.0 $ and~$\beta \approx 3.7$. Dashed line is the  upper bound~$e^{-\left(p+\frac{q}{2}\right)t}\left(e\frac{q}{2}t\right)^j$ of Theorem~\ref{thm:1D_decay}. Here, $p=0.01$, $q=0.1$, and $t = \frac4q$. (B)~Fitted value of~$\alpha$ as a function of~$qt$~(\protect \tikz \protect\draw[fill=mypink] (0,0) -- (0,0.2)--(0.2,0.2)--(0.2,0)--(0,0);). Dashed line is~$(p+\frac{q}{2})t$. (C)~Fitted values of~$\beta$ as a function of~$qt$ (\protect \tikz \protect\draw[fill=mypink] (0,0) -- (0,0.2)--(0.2,0.2)--(0.2,0)--(0,0);). These values lie on the solid line~$0.63qt\approx e\frac{q}{4}t$. Dashed line is the theoretical bound~$e\frac{q}{2}t$.
			}
		\label{fig:decay_simul}
	\end{center}
\end{figure}

The  upper bound in Theorem~\ref{thm:1D_decay} is not sharp.  Numerical simulations show that $f^{\rm 1D}-f^{\mathbb{Z}^+}_{j} \approx e^{-\alpha(t)}\left(\frac{\beta(t)}{j}\right)^j$, see Figure~\ref{fig:decay_simul}A, i.e., that {\em boundary effects do
decay as~$\frac{1}{j^j}$}.  These simulations also show that~$\alpha(t)$ is considerably larger than the theoretical bound~$(p+\frac{q}{2})t$ (Figure~\ref{fig:decay_simul}B), and that~$\beta(t)\approx e\frac{q}{4}t$ (Figure~\ref{fig:decay_simul}C), which is roughly one half the theoretical bound~$e\frac{q}{2}t$.

From Theorem~\ref{thm:1D_decay}, we can obtain a spatial estimate
 that shows that boundary effects
decay exponentially in~$j$,  uniformly for all times:
\begin{corollary}
	\label{cor:two-sided_decay-spatial-only}
 Let $p,q>0$.  Then
	\begin{equation*}
		0< f^{\rm 1D}(t;p,q) - f^{\mathbb{Z}^+}_{j}(t;p,q)< \left(\frac{\frac{q}{2}}{p+\frac{q}{2}}\right)^j, \qquad  0< t <\infty,
		\quad  j \in \mathbb{Z}^+.
	\end{equation*}	
\end{corollary}
\proof{Proof.} See Appendix~\ref{app:1D_decay_1D}. \halmos

Going back to the finite line, the left and right boundaries lower the adoption probability of nodes, as follows:
\begin{lemma}
\label{lem:1D_decay_finite}
 Let $p,q>0$.
 \begin{enumerate}
 	\item Let $t>0$. Then for any $j$ such that $j\ge t\frac{q}{2}$ and $M+1-j \ge t\frac{q}{2}$,
\begin{equation}
\label{eq:1D_decay_finite}
0<f^{\rm 1D}(t;p,q) - f_j^{\left[1,\ldots,M\right]}(t;p,q)
<  e^{-\left(p+\frac{q}{2}\right)t}
\Bigg( \underbrace{\left(\frac{e\frac{q}{2}t}{j}\right)^j}_\textrm{\rm left bdry}+ \underbrace{\left(\frac{e\frac{q}{2}t}{M+1-j}\right)^{M+1-j}}_\textrm{\rm right bdry}
\Bigg).
\end{equation}
  \item For any $ 0< t <\infty$ and  $j \in \cal M$,
  \begin{equation}
  \label{eq:1D_decay_finite-global-in-t}
  	0< f^{\rm 1D}(t;p,q) - f_{j}^{[1,\ldots,M]}(t;p,q)< \underbrace{\left(\frac{\frac{q}{2}}{p+\frac{q}{2}}\right)^j}_\textrm{\rm left bdry}
  	+ \underbrace{\left(\frac{\frac{q}{2}}{p+\frac{q}{2}}\right)^{M+1-j}}_{\rm right~ bdry}.
  \end{equation}
 \end{enumerate}	
\end{lemma}
\proof{Proof.} See Appendix~\ref{app:1D_decay_1D}. \halmos

The first and second terms on the right-hand sides of~\eqref{eq:1D_decay_finite}
and~\eqref{eq:1D_decay_finite-global-in-t} are upper bounds for the effects of the left and right boundaries, respectively.  For a fixed~$t$, each of these effects decays super-exponentially with the distance $j$ and $M+1-j$ from the respective boundary,
respectively, so long that these distances 
are greater than~$\frac{q}{2}t$.
Each of these effects also decays exponentially with the distance from the respective boundary,
   uniformly for all times.

\section{Local Boundary effects in multi dimensions}
\label{sec:higher_dimensions_results}

To extend the analysis of boundary effects to $ D \ge 2$, we consider the discrete Bass model on the
half-space $\mathbb{Z}^+ \times \mathbb{Z}^{D-1} =  [1,\ldots,\infty)\times (-\infty,\ldots,\infty)^{D-1}$.
Let
$$
{\bf j}:=(j_1, {\bf j}_{-1}) \in \mathbb{Z}^+ \times \mathbb{Z}^{D-1},
\qquad {\bf j}_{-1}:=(j_2, \dots, j_D).
$$
Then
\begin{subequations}
	\label{eq:Bass-model-2D+}
	\begin{equation}
		\label{eq:general_initial-2D+}
		X_{{\bf j}}(0)=0, \qquad {{\bf j}} \in \mathbb{Z}^+ \times \mathbb{Z}^{D-1},
	\end{equation}
	and as $ \Delta t \to 0$,
	\begin{equation}
		\label{eq:Bass-model-2D+-dynamics}
		\mathbb{P} (X_{{\bf j}}(t+\Delta t )=1 \, \vert \,  {\bf X}(t))=
		\begin{cases}
			\left(p+\frac{q}{2D}N_{{\bf j}}(t) \right) \Delta t , & {\rm if}\ X_{{\bf j}}(t)=0,
			\\
			\hfill 1,\hfill & {\rm if}\ X_{{\bf j}}(t)=1,
		\end{cases}
		\qquad {\bf j} \in \mathbb{Z}^+ \times \mathbb{Z}^{D-1},
	\end{equation}
	where
	\begin{equation}
		\label{eq:N_j(t)-D+}
		N_{{\bf j}}(t):=  \sum_{i=1}^D \left( X_{{{\bf j}} - \hat{\bf e}_i}(t)
		+X_{{{\bf j}} + \hat{\bf e}_i}(t)
		\right),  \qquad {{\bf j}} \in \mathbb{Z}^+ \times \mathbb{Z}^{D-1},
	\end{equation}
	is the number of adopters connected to~${\bf j}$,
	and 	
	\begin{equation}
		X_{(j_1=0,{\bf j}_{-1})}(t):= 0, \qquad {\bf j}_{-1} \in \mathbb{Z}^{D-1}.
	\end{equation}
	%
\end{subequations}
The analysis in multi-dimensions is harder, since we do not have explicit
expressions such as~\eqref{eq:local_effect-semi-infinite-line}. Nevertheless,
in what follows, we show that all the results in 1D  extend naturally to multi-dimensions.

\subsection{Maximal local boundary effect}
\label{sec:sub_local_D}


Denote
the adoption probability of node~${\bf j} = (j_1,{\bf j}_{-1})$
in the discrete Bass model~\eqref{eq:Bass-model-2D+} 
by~$f^{\mathbb{Z}^+ \times \mathbb{Z}^{D-1}}_{(j_1,{\bf j}_{-1})}$.

\begin{lemma}
	\label{lem:f^N*Z_j1_j2}
	Let $p,q,t>0$. Then
	$f^{\mathbb{Z}^+ \times \mathbb{Z}^{D-1}}_{(j_1,{\bf j}_{-1})}(t;p,q)$
	is independent of~${\bf j}_{-1}$, and
	is monotonically  increasing in~$j_1$.
\end{lemma}
\proof{Proof.} See Appendix~\ref{app:f^N*Z_j1_j2}. \halmos

The monotonicity in~$j_1$ implies that
the maximal local boundary effect is obtained
by comparing the two extreme cases of boundary and central nodes, i.e.,
$$
f^{\mathbb{Z}^+ \times \mathbb{Z}^{D-1}}_{\rm bdry}:=
f^{\mathbb{Z}^+ \times \mathbb{Z}^{D-1}}_{(j_1=1,{\bf j}_{-1})},
\qquad
f^{\mathbb{Z}^+ \times \mathbb{Z}^{D-1}}_{\rm central }:=\lim_{j_1 \to \infty}
f^{\mathbb{Z}^+ \times \mathbb{Z}^{D-1}}_{(j_1,{\bf j}_{-1})}.
$$

As in  1D, see Lemma~\ref{lem:local_effect-semi-infinite-line}, the effect of the boundary disappears as $j_1 \to \infty$:
\begin{lemma}
	\label{lem:f_D_central=f_D}
		Let $p,q>0$. Then
	$$
	f^{\mathbb{Z}^+ \times \mathbb{Z}^{D-1}}_{\rm central }(t;p,q)
	=f^{\rm D}(t;p,q),
	$$
	where $f^{\rm D}$ is the expected adoption level in the discrete Bass model~\eqref{eqs:Bass-general_D} on~$\mathbb{Z}^D$.
\end{lemma}
\proof{Proof.}
	This follows from Lemma~\ref{lem:Dtwo-sided} below.
\halmos

As in 1D (Lemma~\ref{lem:R-boundary-1D}),
the adoption probability of boundary nodes is lower than of central nodes:
\begin{corollary}
	\label{cor:f_2D_bdry<f_2D_central}
	For any $p,q>0$,
	\begin{equation}
		\label{eq:f_D_bdry<f_D_central}
		f^{\mathbb{Z}^+ \times \mathbb{Z}^{D-1}}_{\rm bdry}(t;p,q)<f^{\mathbb{Z}^+ \times \mathbb{Z}^{D-1}}_{\rm central }(t;p,q), \qquad t>0.
	\end{equation}
\end{corollary}
\proof{Proof.}
	This follows from Lemma~\ref{lem:f^N*Z_j1_j2}.
\halmos

\begin{figure}[ht!]
	\begin{center}
		\scalebox{0.8}{\includegraphics{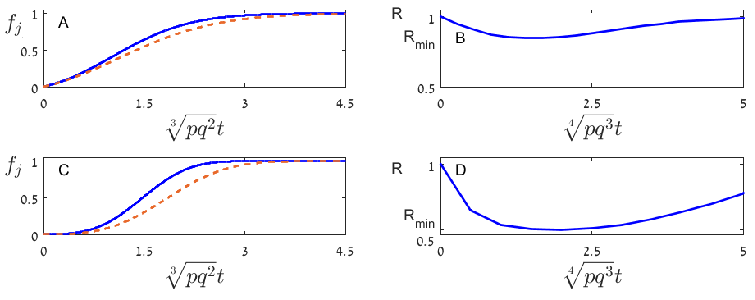}}
		\caption{(A)~The adoption probability of central ($f^{\mathbb{Z}^+ \times \mathbb{Z}}_{\rm central}$, blue solid) and  boundary ($f^{\mathbb{Z}^+ \times \mathbb{Z}}_{\rm bdry}$, orange dashes) nodes on~$\mathbb{Z}^+ \times \mathbb{Z}$,
			as a  function of~$\sqrt[3]{pq^2}t$.   Here~$\frac{q}{p} = 10$. (B)~The ratio~$R:=f^{\mathbb{Z}^+ \times \mathbb{Z}}_{\rm bdry}/f^{\mathbb{Z}^+ \times \mathbb{Z}}_{\rm central}$ as a function of~$\sqrt[4]{pq^{3}}t$. (C) and (D):~Same as~(A) and~(B) with~$\frac{q}{p} = 10^6$.
			The curves are obtained from simulations of the discrete Bass model~\eqref{eq:Bass-model-2D+}.
		}
		\label{fig:2D_local_effect}
	\end{center}
\end{figure}

Relation~\eqref{eq:f_D_bdry<f_D_central} is illustrated numerically
in Figure~\ref{fig:2D_local_effect}, for $D=2$.
These simulations also suggest that, as in 1D (see Lemma~\ref{lem:R-boundary-1D}),
the adoption probability of boundary nodes is at least half  that of the central nodes (Figures~\ref{fig:2D_local_effect}B and~\ref{fig:2D_local_effect}D), i.e.,
that
$$
		\frac12 f^{\mathbb{Z}^+ \times \mathbb{Z}^{D-1}}_{\rm  central}(t)
<
	f^{\mathbb{Z}^+ \times \mathbb{Z}^{D-1}}_{\rm  bdry}(t)
\qquad t>0.
$$
In 1D we proved this inequality using the explicit expression
$f^{\mathbb{Z}^+}_{\rm bdry}(t;p,q)=f^{\rm 1D}\left(t;p,\frac{q}{2}\right)$,
see~\eqref{eq:local_effect-semi-infinite-line}. There is no similar explicit expression for~$f^{\mathbb{Z}^+ \times \mathbb{Z}^{D-1}}_{\rm bdry}$
(the $D$-dimensional analog of $f^{\rm 1D}\left(t;p,\frac{q}{2}\right)$
is smaller than $f^{\mathbb{Z}^+ \times \mathbb{Z}^{D-1}}_{\rm bdry}$,
see Appendix~\ref{app:f^D_bdry-lower-bound}).
Therefore, we only provide an informal argument in support of the above inequality, as follows.
Any node  adopts either due to external influences, or due to a series of internal adoption events that started from a different node which adopted externally. For any such {\em adoption path} on the half-space grid $\mathbb{Z}^+ \times \mathbb{Z}^{D-1}$ that ends in a boundary node, there is also an identical adoption path on the full-space grid $\mathbb{Z}^{D}$ which is symmetric with respect to the boundary, and has exactly the same probability to occur. Therefore, the overall adoption probability  on the full-space domain is higher than that for a boundary node, but less than twice that for a boundary node.

\subsection{Spatial decay of boundary effects}

As in~1D, for any fixed positive time, {\em the effect of the boundary in~$\mathbb{Z}^+ \times \mathbb{Z}^{D-1}$
	decays at the super-exponential rate of~$\frac{1}{j_1^{j_1}}$}:
%
%
%
\begin{lemma}
	\label{lem:Dtwo-sided}
	Consider the discrete Bass
	model~\eqref{eq:Bass-model-2D+} on~$\mathbb{Z}^+ \times \mathbb{Z}^{D-1}$. Let $p,q>0$.
	Then  for any $t>0$,
	\begin{equation}
		\label{eq:Dtwo-sided}
		0<f^{\rm D}(t;p,q) - f^{\mathbb{Z}^+ \times \mathbb{Z}^{D-1}}_{(j_1,{\bf j}_{-1})}(t;p,q) < 
		e^{-\left(p+\frac{q}{2D}\right) t}  e^{e\frac{D-1}{D}qt}
		\left(\frac{e\frac{q}{2D}t}{j_1}\right)^{j_1}, \qquad j_1 >\frac{q}{2D}t.
	\end{equation}
	\end{lemma}
	\proof{Proof.}
		See Section~\ref{app:2Dtwo-sided}.
	\halmos

	The upper bound in~\eqref{eq:Dtwo-sided} only depends on~$j_1$, since  $f^{\mathbb{Z}^+ \times \mathbb{Z}^{D-1}}_{(j_1,{\bf j}_{-1})}$ is independent of~${\bf j}_{-1}$ (Lemma~\ref{lem:f^N*Z_j1_j2}).
	It has the undesired property that it increases with~$t$ if $\frac{2e(D-1)-1}{2D}q>p$, and so is much less tight than in 1D (Theorem~\ref{thm:1D_decay}), where it decays with~$t$ for any~$p,q>0$ (this is because~\eqref{eq:Dtwo-sided} is derived by summing the individual upper bounds of all possible 1D paths, without discounting for the large overlap between the paths, see~\eqref{eq:fD_suf_cond-boundary}).
	 Nevertheless, the upper bound~\eqref{eq:Dtwo-sided} does capture  the~$\frac{1}{j_1^{j_1}}$ decay as~$j_1 \to \infty$ at any fixed time.
	
		We can extend Lemma~\ref{lem:Dtwo-sided} to finite domains:
	\begin{lemma}
		\label{lem:finite_D_decay}
		Consider the discrete Bass
		model~\eqref{eq:two_sided_line_all-isotropic} on the $D$-dimensional box~$B_D:=[1,\ldots, M_1]^D$.
		Then boundary effects on the adoption probability of node~${\bf j}$ decay
		exponentially with the distances from the boundaries, i.e.,
		\begin{equation}
			\label{eq:finite_D_decay}
			\begin{aligned}
				0<&f^{\rm D}(t;p,q)-f^{B_D}_{\bf j}(t;p,q)
				<		e^{-\left(p+\frac{q}{2D}\right) t}  e^{e\frac{D-1}{D}qt}
				\left[\sum_{i=1}^D \left(\frac{e\frac{q}{2D}t}{j_i}\right)^{j_i}
				+\left(\frac{e\frac{q}{4}t}{M_1-j_i}\right)^{M_1-j_i}
				\right],
			\end{aligned}
		\end{equation}
		for~$\frac{q}{4}t \leq j_i \leq M_1-\frac{q}{2D}t$ , where $i=1, \dots, D$.
	\end{lemma}
	\proof{Proof}
		See Section~\ref{app:2Dtwo-sided}.
	\halmos

\section{Global boundary effects}
\label{sec:global_effect_high_dim}

In Sections~\ref{sec:local_effect_1D} and~\ref{sec:higher_dimensions_results}
we saw that boundary effects decay exponentially
with the distance from the boundary. Therefore, on
the $D$-dimensional cube~$B_D:=[1,\ldots, M_1]^D$,
there are, roughly speaking, $O(M_1^{D-1})$~nodes for which boundary effects are not exponentially small. Since the expected adoption level is averaged over $M = M_1^D$~nodes, the global effect of boundaries should be~$O(\frac1{M_1}) = O(\frac1{M^{1/D}})$.
Indeed, we have
\begin{theorem}
	\label{thm:global_effect2-boundary-2D}
	Let $f^{B_D}$ denote  the expected adoption level in the discrete Bass
	model~\eqref{eqs:Bass-general_D} on~$B_D$, see~\eqref{eq:f^B_D-def}. Then
	\label{lem:f_D_global}
	\begin{equation}
		\label{eq:f_D_global}
		0<f^{\rm D}(t;p,q)-f^{B_D}(t;p,q) = O\left(\frac{1}{M^{1/D}}\right), \qquad M\rightarrow \infty,
	\end{equation}
where $f^{D}$ is  the expected adoption level in the discrete Bass
model~\eqref{eqs:Bass-general_D} on~$\mathbb{Z}^D$.
\end{theorem}
\proof{Proof} See Appendix~\ref{app:global_effect2-boundary-2D}. \halmos

\begin{figure}[ht!]
	\begin{center}
		\scalebox{0.8}{\includegraphics{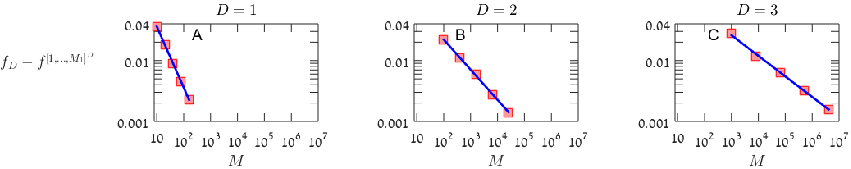}}
		\caption{ $f^{\rm D}-f^{B_D}$ as a function of~$M = M_1^D$, on a log-log scale (\protect \tikz \protect\draw[fill=myred] (0,0) -- (0,0.2)--(0.2,0.2)--(0.2,0)--(0,0);). Here~$\frac{q}{p}=10$ and~$t=\frac{20}{q}$.
			Solid line is the fitted curve~$\log(f^{\rm D}-f^{[1, \dots, M_1]^D}) \sim \alpha_D + \beta_D \log{M}$. (A)~$D = 1$, $\alpha_1 \sim -1.02$, $\beta_1 \sim -1.00$. (B)~$D = 2$, $\alpha_2 \sim -1.54$, $\beta_2 \sim -0.49$. (C)~$D=3$, $\alpha_3 \sim -1.32$, $\beta_3 \sim -0.34$.
			.
		}
		\label{fig:global_effect_2D}
	\end{center}
\end{figure}

The theoretical $O\left(\frac{1}{M^{1/D}}\right)$ reduction in~$f$ by the boundaries  is confirmed numerically in Figure~\ref{fig:global_effect_2D}
for $D=1$, $2$, and~$3$, where  $f^{\rm D}-f^{B_D} \sim \frac{c_D}{M^{\beta_D}}$
and 
$\beta_D \approx \frac1D$ with
a two-digit accuracy. 

\subsection{Global  effect of boundaries in 1D}
\label{sec:global-effect-1D}


In 1D, we can obtain an {\em explicit expression} for the {\em leading-order global effect} of the boundary:

\begin{theorem}
	\label{thm:global_effect2-boundary}
	Let $f^{[1,\ldots,M]}$ denote  the expected adoption level in the discrete Bass
	model~\eqref{eq:two_sided_line_all-isotropic} on~$[1, \dots,M]$. Then
	\begin{equation}
		\label{eq:global_effect2}
		f^{[1,\ldots,M]}(t;p,q)-f^{\rm 1D}(t;p,q) \sim -\frac{2\psi(t;p,q)}{M}, \qquad M\rightarrow \infty,
	\end{equation}
	where $f^{\rm 1D}$ is the adoption level on the infinite line,
	see~\eqref{eq:f_1D=lim}, and
	\begin{equation}
		\label{eq:global_effect_psi-1D}
		\psi(t;p,q) := \sum_{j=1}^{\infty}\left(f^{\rm 1D}(t;p,q)-f_j^{\mathbb{Z}^+}(t;p,q)\right).
	\end{equation}
\end{theorem}
\proof{Proof.} See Appendix~\ref{app:global_effect2-boundary}. \halmos


The function~$\psi$ has the following properties:

\begin{lemma}
	\label{lem:psi-1D}
	Let $\psi$ be given by~\eqref{eq:global_effect_psi-1D}. Then
	\begin{enumerate}
		\item $0<\psi(t;p,q) < \frac{q}{2p}<\infty$.
		\item $\lim_{t \to 0} \psi(t;p,q) = \lim_{t \to \infty}\psi(t;p,q) = 0$.
		\item $\lim_{\frac{q}{p} \to 0}\psi(t;p,q) = 0$, uniformly in~$t$.
	\end{enumerate}
\end{lemma}
\proof{Proof.} See Appendix~\ref{app:psi-1D}. \halmos
	%
	%

\begin{figure}[ht!]	
	\begin{center}
		\scalebox{0.8}{\includegraphics{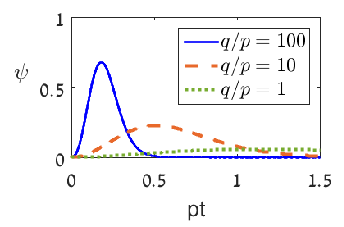}}
		\caption{$\psi$
			as a function of time, for~$\frac{q}{p}=100$ (blue solid), $\frac{q}{p}=10$ (orange dashes), and~$\frac{q}{p}=1$ (green dots).}
		\label{fig:psi_1D_line}
	\end{center}			
\end{figure}
These properties are illustrated in Figure~\ref{fig:psi_1D_line}.
Thus, $\psi$ is positive since the boundary slows down the diffusion,
there are no boundary effects as~$t \to 0$ since everyone is a nonadopter,
as~$t \to \infty$ since everyone is an adopter, and as $\frac{q}{p} \to 0$
since internal influences become negligible, hence also boundary effects.

\begin{remark}
The function $\psi$ is also the $\frac{1}{M}$ leading-oder global effect of an {\em internal boundary} in~1D,
see Lemma~\ref{lem:h_M-1D}.
\end{remark}

\subsubsection{Boundary versus network-size effects}

The difference between $f$ on finite and infinite lines can be written as, see~\eqref{eq:f_1D=lim},
\begin{equation}
	\label{eq:f^[1,ldots,M]-f^1D-long}
	\begin{aligned}
		f^{\left[1,\ldots,M\right]}-f^{\rm 1D}
		=\underbrace{f^{\left[1,\ldots,M\right]}-f^{\rm circle}(\cdot,M)}_\textrm{boundary effect}
		+
		\underbrace{f^{\rm circle}(\cdot,M)-\lim_{M\rightarrow \infty}f^{\rm circle}(\cdot,M)}_\textrm{network-size effect}.
	\end{aligned}
\end{equation}
Therefore, there are two sources for the $O(\frac1M)$ difference between~$f^{\rm 1D}$ and~$f^{\left[1,\ldots,M\right]}$ in Theorem~\ref{thm:global_effect2-boundary}:
\begin{enumerate}
	\item Existence/non-existence  of a boundary.
	\item Difference in network size.
\end{enumerate}
Comparing the diffusion on a finite and infinite circles
(Lemma~\ref{lem:circle_diff}) shows that {\em the effect of network-size on~$f$ is exponentially small} in~$M$, i.e.,
\begin{equation}
	\label{eq:f_circle_diff-B}
	0<\lim_{M\rightarrow \infty}f^{\rm circle}(t;p,q,M)-f^{\rm circle}(t;p,q,M)< e^{-(p+q)t}\left(\frac{eqt}{M}\right)^M, \qquad  M  \to \infty.
\end{equation}
Therefore, by~\eqref{eq:f^[1,ldots,M]-f^1D-long} and~\eqref{eq:f_circle_diff-B},
{\em the effect of the boundary on~$f$
	is~$O(\frac1M)$},
i.e.,
\begin{equation*}
	f^{\left[1,\ldots,M\right]}(t;p,q)-f^{\rm circle}(t;p,q,M) \sim - \frac{2\psi(t;p,q)}{M}, \qquad M\rightarrow \infty.
\end{equation*}
In particular, 
{\em the~$O(\frac1M)$ difference in $f$  between finite and infinite lines  in Theorem~\ref{thm:global_effect2-boundary} is due to the boundary, and not to the difference in network size.}

\section{Empirical Evidence of Boundary Effects} 
\label{sec:empirical}

The theory laid out thus far has direct implications for the patterns of diffusion of new technologies. In this section, we use the example of residential rooftop solar photovoltaic systems to illustrate the power of the theory by exploring spatial data on adoption patterns. We consider the local effect of external boundaries, which was analyzed in Section~\ref{sec:higher_dimensions_results}, and explore whether there is empirical evidence that the adoption of solar by residential units near the municipality boundary is substantially lower than that of centrally-located units.

Our empirical example is solar adoption in Connecticut, which has a vibrant market for rooftop solar. From the start of the solar market around~2005, when there are no installations in Connecticut, to December~2019, over~290 MW of rooftop solar capacity has been installed, consisting of over~35,000 installations. In our analysis, we use administrative data from the Connecticut Green Bank, which collects data on nearly every solar installation in the state (there are two small municipal utilities and these utilities run their own programs, so we do not observe solar adoptions in these municipalities). These data, obtained under a non-disclosure agreement, include the date the contract was signed to install solar, the date the installation was completed, the size of the system, and the address of the installation. We observe all solar systems installed from~2005 through~2019, but for this analysis we focus on the period after~2012 when there were a sufficient number of installations for a meaningful statistical analysis.

The year~2012 is also useful because it is when the set of solar campaigns that we exploit for our analysis began. This set of campaigns is extremely useful because it provides a substantial boost to solar installations, thus allowing for sufficiently powered statistical analyses. These grassroots campaigns, called {\em Solarize campaigns}, aimed to encourage solar adoption by fostering word-of-mouth and were run by a non-profit in cooperation with the state agency, the Connecticut Green Bank. The campaigns involved a limited time frame of roughly~20 weeks, municipality-chosen installers, volunteer community members who served as solar ambassadors to tell their friends about solar, mailings to all households in the municipalities, local media attention, and multiple events to inform residents. \cite{gillingham2021social} showed that the rate of solar adoptions increases by over~1,000\% during the time frame of the campaigns, and the elevated rate of adoption continued after the campaigns relative to control municipalities. While we avoid the actual campaign period itself to avoid any influence of the marketing itself, the boost in adoption in the year after provides substantial statistical power for an analysis of solar installations in different locations within a municipality. The campaigns were run over several rounds, with anywhere from 6 to 18 municipalities in each round from~2012 through~2019, for a total of~51 municipality-level campaigns analyzed in this analysis. Importantly, {\em the campaigns were run uniformly at the municipality level} and there were no central foci for the campaigns. For example, mailers went to all residents, events were held in many locations around the municipality, and ambassadors were encouraged to reach anyone and everyone in the municipality.

For our analysis we use ArcGIS and Python to geocode all of the addresses to obtain the latitude and longitude. We count the number of solar installations within a small distance (e.g., one mile) of the municipality boundary. We call this region within the municipality close to the border the {\em boundary-buffer zone}. We also count of the number of owner-occupied homes within the same distance to calculate the {\em market share} (i.e., fraction) of owner-occupied homes that had an installation during the one year just after the Solarize.  We then compare the number of solar installations in the boundary-buffer zone to the rest of the municipality, which we call the {\em inner-core zone}. We focus on the adoption process on one year after each of the Solarize campaigns ended to assure that the campaigns are fully over and there is sufficient time for the campaigns to exogenously lead to a large increase in adoptions. All municipalities with Solarize campaigns since 2012 were included, and if there were multiple Solarize campaigns by a municipality, only the first one is included.

In this analysis, the use of Solarize campaigns is valuable for three reasons. First, we need enough solar adoptions in each municipality for a sufficiently strong signal-to-noise ratio. For many municipalities that do not have Solarize campaigns, we would not have enough rooftop solar adoptions to test our theory with reasonable statistical power. Second, the Solarize campaigns provide a nice uniform exogenous shock in rooftop solar installations across the entire municipality that could rule out some confounders, such as solar firms not working in some parts of the municipality. Third, using Solarize campaigns illustrates that  boundary effects also occur with non-physical boundaries (i.e., the boundary between two bordering municipalities; one with a Solarize campaign and the second without.)

We estimate the following model specification in a fixed-effects regression where solar adoption in municipality $i$ and zone category $j \in \{\text{boundary buffer}, \text{inner core}\}$ is given as follows:
\begin{equation*}
(\mathrm{installed} \ \%)_{i,j} = \beta \, \mathbbm{1}^{\mathrm{boundary}}_{j} + \mathbf{X}_{i,j} \gamma + \mu_i + \epsilon_{i,j}.
\end{equation*}
Here~$(\mathrm{installed} \ \%)_{ij}$ is the market share of rooftop solar (fraction of owner-occupied homes) installed in the one year after the end of each municipality's Solarize campaign ended,  $\mathbbm{1}^{\mathrm{boundary}}$ is an indicator variable for the boundary buffer zone, $\mathbf{X}_{ij}$ is a matrix of control variables, which include median household income, median number of rooms in owner-occupied households, median home values, and population density, and $\mu_i$ are fixed effects for municipalities. Therefore, $\beta$ quantifies the reduction in the adoption of residential rooftop solar in the boundary-buffer.

In this analysis, the inner-core zone acts as a control group for the boundary buffer after including the controls. The key assumption for our analysis to be quantifying a causal effect of boundaries on rooftop solar adoption is that the households in the inner core buffer zone serve as a valid control group after including the controls. This seems very plausible because the inner-core zone is in exactly the same municipality as the boundary buffer, the Solarize campaigns are run uniformly across the municipalities, and we have controlled for any differences in homes or income. The summary statistics for the data are given in Table~\ref{tab:sumstatsinternal}.

The main finding from this analysis is a very strong significant reduction in installations per owner-occupied household in the boundary buffer zones. Before even getting to the regression, the basic summary stats bear this out. The mean percentage of owner-occupied dwellings that have solar installations during the Solarize campaign period is 1.37\% 
in the inner core and 0.97\% in the boundary buffer zone, a difference of 0.40 percentage points 
(p-value=0.016 for t-test of equality of means). Adding controls in a regression framework makes little difference.

Our preferred estimation that includes all of the controls results yields 
$$\beta = -0.41.
$$
 This indicates that the probability of adoption is 0.41 percentage points less in the one-mile boundary buffer zone than in the inner-core zone, which amounts to reducing the probability of adoption by 30\% (0.413/1.37). This coefficient is significant (p-value=0.000). The full estimation results can be found in Table~\ref{tab:internal}. To further confirm these results, we also run a robustness check by examining non-Solarize municipalities and the first half of the Solarize campaigns. The results are similar, although the effects are smaller (as would be expected with less word-of-mouth about solar) and most statistical significance is lost, as can be seen in Table~\ref{tab:empiricalrobust}. These municipalities were chosen because they never had a Solarize campaigns and were not adjacent to other Solarize campaigns, but we analyzed a period that aligned with the other Solarize campaigns. The main takeaway is that the boundary zones very clearly have a lower probability of adoption per owner-occupied household than the internal zones, likely due to fewer social interactions or peer effects in the boundary zone. We view this as solid evidence from the data that boundaries affect the diffusion of solar, matching our theoretical findings in Section~\ref{sec:higher_dimensions_results}.

To the best of our knowledge, this is the first empirical observation of boundary effects in diffusion of new products.  From an empirical methodological perspective, this is the first paper we are aware of that develops a geographical information systems (GIS) buffer approach to examine boundary effects in real-world data, as well as leverage marketing campaigns to examine boundary effects.

\section{Discussion}
  \label{sec:discussion}



\subsection{Model Robustness}

In this study we used the Bass model on a 2D grid to study the effect of boundaries on diffusion of new products such as PV solar, for which peer effects are primarily exerted by geographical neighbors.
To gain confidence in the robustness of our results, we should test whether they remain valid on
networks with a more realistic  structure, namely,
on a 2D Cartesian network which is perturbed by a small fraction of random long-range connections.  The motivation for this robustness test is that
the addition of a small
fraction of long-range connections to a Cartesian network in epidemiological models can lead to a dramatic speedup in the spread of epidemics~\citep{watts1998collective}.

In the discrete Bass model for the diffusion of new products, however,
the addition of a small fraction of random long-range connections
has a minor effect of the diffusion \citep{OR-10,Bass-SIR-model-16}.
Briefly, the difference between epidemiological models and the Bass model is that an
epidemic starts from a few
infected individuals at $t = 0$ (“patients zero”),
and then progresses only through internal influences (i.e., is modeled using discrete SI/SIR models in which $p=0$). In
that case, the key network property that determines the diffusion speed is the average distance of nodes from
{\em patients zero}, which is highly sensitive to the addition
of long-range connections.  In diffusion of new products, however,
there is an ongoing random creation of adopters by external influences.
Once an external adopter is created, it expands into a {\em cluster of adopters} through internal adoptions. Therefore, the diffusion speed is
determined by the growth rate of clusters,
which depends on local properties of the network, but not on
global properties such as the average distance from {\em patients zero}.

\begin{figure}[ht!]	
	\begin{center}
		\scalebox{0.6}{\includegraphics{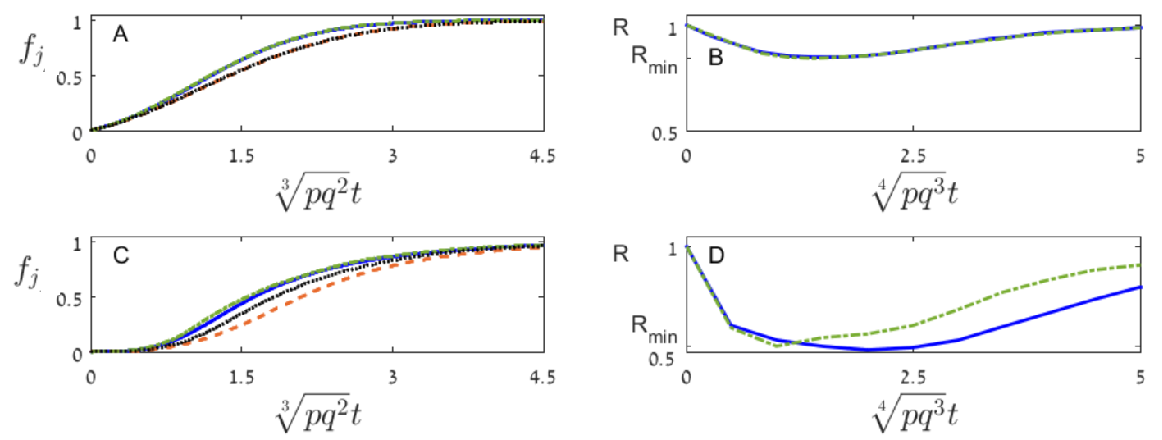}}
		\caption{(A)~The adoption probability of central ($f^{\mathbb{Z}^+ \times \mathbb{Z}}_{\rm central}$, green dash-dots) and  boundary ($f^{\mathbb{Z}^+ \times \mathbb{Z}}_{\rm bdry}$, black dots) nodes on~$\mathbb{Z}^+ \times \mathbb{Z}$,
			as a  function of~$\sqrt[3]{pq^2}t$.   Here~$\frac{q}{p} = 10$. (B)~The ratio~$R:=f^{\mathbb{Z}^+ \times \mathbb{Z}}_{\rm bdry}/f^{\mathbb{Z}^+ \times \mathbb{Z}}_{\rm central}$ as a function of~$\sqrt[4]{pq^{3}}t$ (green dash-dots). (C) and (D):~Same as~(A) and~(B) with~$\frac{q}{p} = 10^6$.
			The curves are obtained from simulations of the discrete Bass model~\eqref{eq:Bass-model-2D+}.
			The curves from Figure~\ref{fig:2D_local_effect} are also showed for comparison. }
		\label{fig:local_long_range_random_and_no_random}
	\end{center}
\end{figure}

To test this numerically, in Figure~\ref{fig:local_long_range_random_and_no_random}  we repeat the simulations  of Figure~\ref{fig:2D_local_effect}  on local  boundary effects
in 2D, this time adding 5\% random long-range connections
(i.e., any two nodes have probability $\frac{5\%}{M-1}$ to be connected by an edge with weight $\frac{q}{2D}$).
The curves with and without the random long-range connections
are nearly indistinguishable. The only case where the agreement is only qualitative is for
boundary nodes, the extreme value of~$\frac{q}{p} = 10^6$ where there are so few external adoptions,
and the addition of a relatively large (5\%) fraction of random long-range connections.
Even in that extreme case, however, the qualitative predictions that the adoption of boundary nodes is slower than that of central nodes,
but is at least half that of central nodes, remains valid in the presence of random long-range connections.

\begin{figure}[ht!]	
	\begin{center}
		\scalebox{0.8}{\includegraphics{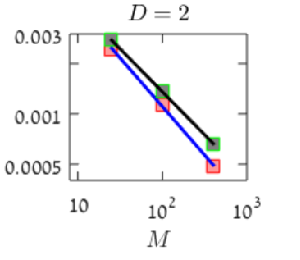}}
		\caption{$f^{\rm 2D}(t;p,q)-f^{B_{\rm 2D}}(t;p,q)$ as a function of~$M = M_1^2$ and on a log-log scale,
	    for 2D grids without (\protect \tikz \protect\draw[fill=myred] (0,0) -- (0,0.2)--(0.2,0.2)--(0.2,0)--(0,0);) and with (\protect \tikz \protect\draw[draw=mygreen, fill=mygrey] (0,0) -- (0,0.2)--(0.2,0.2)--(0.2,0)--(0,0);) $5\%$ random long-range connections.
			The fitted solid lines are $\log(f^{\rm 2D}-f^{B_{\rm 2D}}) \sim \alpha_2 + \beta_2 \log{M}$, with $(\alpha_2,\beta_2)=-(0.53,4.13)$ and
						with~$(\alpha_2,\beta_2)=-(0.51,4.24)$, respectively.
	   Here~$\frac{q}{p}=10$ and~$t=\frac{5}{q}$.}
		\label{fig:global_long_range}
	\end{center}			
\end{figure}

In Figure~\ref{fig:global_long_range}  we compute the global boundary effect on the aggregate adoption level on a 2D~grid, with and without 5\% random long-range connections.
In both cases, the theoretical $O\left(\frac{1}{M^{1/2}}\right)$ reduction in~$f$ by the boundaries  is confirmed numerically, as  $\alpha_2 \approx \frac12$. 
In particular, it remains valid under the addition of random long-range connections.
{\em Together, Figures~\ref{fig:local_long_range_random_and_no_random} and~\ref{fig:global_long_range} demonstrate the robustness of the 2D Bass model under the addition of a small fraction of random long-range connections.}
			
The reduced adoption probability for boundary nodes does not require the precise adoption mechanism of the Bass model
which was used in this study. Indeed, consider any stochastic adoption mechanism on a 2D Cartesian grid,
where any node can adopt externally, and adopting nodes can influence their peers to adopt. Then the adoption by any node can be traced backward to an adoption path that
starts from some node that adopted externally, followed by a series of peer to peer contagion events.
From any node that adopts externally, there exist adoption paths that reach all interior and boundary nodes.
The average length of the adoption paths that reach a boundary node, however, is substantially longer than
for interior nodes. In addition, since a longer adoption path requires more contagion events, its probability is lower.
Therefore, the overall adoption probability of boundary nodes is smaller.

The discrete 2D Bass model does not take into account the feature that peer effects of solar installations on their neighbors decay after several months~\citep{Graziano-15}. A more comprehensive model for the diffusion of solar is the discrete Bass-SIR model~\citep{Bass-SIR-model-16,Bass-SIR-analysis-17}, in which adopters stop influencing their peers after some time. The 2D~Bass-SIR model can also be used for the spread of epidemics, in which the probability of an infection decays with distance and with time, and for which
there is a continuous influx (immigration) of infected individuals into the domain, such as Covid~$19$. Boundary effects in the spatial-temporal dynamics of such epidemics may occur, for example, at the boundary between two U.S. states that adopt different epidemiological policies. Our results are expected to remain valid for the Bass-SIR model, namely, that individuals near a boundary are less likely to get infected, and that boundary effects on the infection probability of nodes decay exponentially with the distance from the boundary. A systematic analysis of that model, however, is left for a future study.

\subsection{Final remarks}

This study draws attention to the role that boundaries play in the spread of solar. In particular, it shows that the distance of an adopter from the town boundary is important for its expected influence on other non-adopters. This observation suggests that if we have scarce resources that enable us to focus on limited number of residential units (via financial incentives or direct marketing), we should focus on central residential units, rather than on the boundary units.

At a more fundamental level, the striking agreement between the theoretical prediction that boundary nodes adopt considerably slower than central nodes, and our empirical findings,
provides further support to the robustness of the discrete Bass model on 2D Cartesian grids, which has important managerial implications in that is suggests that the Bass model may have further usefulness beyond what was previously understood. All models are simply representations of reality and future work could explore the conditions under which the Bass model works the best for modeling the diffusion of solar, as well as other related pro-social products.


\begin{APPENDICES}

\section{Proof of Lemma~\ref{lem:local_effect-semi-infinite-line}}
\label{app:local_effect}

The adoption probabilities of the two boundary nodes are~\citep{Bass-boundary-18}
\begin{equation}
	\label{eq:f_1=f_M}
	f_{j=1}^{\left[1,\ldots,M\right]}(t;p,q)=f_{j=M}^{\left[1,\ldots,M\right]}(t;p,q)=f^{\rm circle}\left(t;p,\frac{q}{2},M \right),
\end{equation}
where $f^{\rm circle}$
is the expected adoption level in a circle with~$M$ nodes (see Section~\ref{sec:circle}).
%
Letting~$M\rightarrow \infty$ and using~\eqref{eq:f_1D=lim} yields the left relation of~\eqref{eq:local_effect-semi-infinite-line}.
Let $[S^{\rm circle}]:=1-f^{\rm circle}$ and ~$[S_j^{[1,\ldots,M]}]:= 1-f_j^{[1,\ldots,M]}$ be the non-adoption probabilities of nodes on the circle and on the line, respectively.
\cite{Bass-networkeffect-20} showed that
\begin{equation}
	\label{eq:S_j_funnel}
	[S_j^{[1,\ldots,M]}](t;p,q)  = e^{p t}[S^{\rm circle}]\left(t;p,\frac{q}{2},j \right)
	[S^{\rm circle}]\left(t;p,\frac{q}{2},M+1-j \right), \qquad j=1,\ldots,M.
\end{equation}

Letting $ M \to \infty$ and using~\eqref{eq:f_1D=lim} gives
\begin{equation}
\label{eq:S_j_semi_infinite}
\begin{aligned}
[S^{\mathbb{Z}^+}_j](t;p,q)  & =   \lim_{M \to \infty}[S^{[1, \dots, M]}_j](t;p,q)  =   \lim_{M \to \infty}e^{p t}[S^{\rm circle}]\left(t;p,\frac{q}{2},j \right)
[S^{\rm circle}]\left(t;p,\frac{q}{2},M+1-j \right)
\\ &=
 e^{p t}[S^{\rm circle}]\left(t;p,\frac{q}{2},j \right)
[S^{\rm 1D}]\left(t;p,\frac{q}{2} \right),
\end{aligned}
\end{equation}
where $[S_j^{\mathbb{Z}^+}]:=1-f_{j}^{\mathbb{Z}^+}$.
Letting~$j\rightarrow \infty$ and using~\eqref{eq:f_1D=lim} again gives~$[S^{\mathbb{Z}^+}_{\rm central}](t;p,q) = e^{pt} [S^{\rm 1D}]^2(t;p,\frac{q}{2})$.
Since
\begin{equation}
\label{eq:S_1D=S_1D^2}
e^{pt}[S^{\rm 1D}]^2\left(t;p,\frac{q}{2} \right) =  e^{pt}\left(e^{-\left(p+\frac{q}{2}\right)t+\frac{q}{2p}\left(1-e^{-pt}\right)}\right)^2 = e^{-(p+q)t+\frac{q}{p}\left(1-e^{-pt}\right)} = [S^{\rm 1D}](t;p,q),
\end{equation}
see~\eqref{eq:f_1D=lim}, then~$[S^{\mathbb{Z}^+}_{\rm central}](t;p,q)=  [S^{\rm 1D}](t;p,q)$. Hence, the right relation of~\eqref{eq:local_effect-semi-infinite-line} follows.

\section{Proof of Lemma~\ref{lem:R-boundary-1D}}
\label{app:R-boundary-1D}

	The right inequality 
follows from  Lemma~\ref{lem:local_effect-semi-infinite-line},
since~$f^{\rm 1D}(t;p,q)$ is  monotonically increasing in~$q$, see~\eqref{eq:f_1D=lim}.
To prove the left inequality, we first note that for~$t>0$,
$$1-2[S^{\rm 1D}]\left(t;p,\frac{q}{2}\right)+e^{pt}[S^{\rm 1D}]^2 \left(t;p,\frac{q}{2}\right)=\left(1-[S^{\rm 1D}]\left(t;p,\frac{q}{2}\right)\right)^2+[S^{\rm 1D}]^2 \left(t;p,\frac{q}{2}\right)\left(e^{pt}-1\right)>0.
$$
Hence,
$
1-e^{pt}[S^{\rm 1D}]^2 \left(t;p,\frac{q}{2}\right) < 2\left(1-[S^{\rm 1D}]\left(t;p,\frac{q}{2}\right)\right)$.
Therefore, by~\eqref{eq:S_1D=S_1D^2},
$$
\frac{f^{\rm 1D}(t;p,q)}{f^{\rm 1D}\left(t;p,\frac{q}{2}\right)}=\frac{1-[S^{\rm 1D}](t;p,q)}{1-[S^{\rm 1D}]\left(t;p,\frac{q}{2}\right)}
= \frac{1-e^{pt}
	[S^{\rm 1D}]^2 \left(t;p,\frac{q}{2}\right)}{1-[S^{\rm 1D}]\left(t;p,\frac{q}{2} \right)}<2, 
$$
which, together with~\eqref{eq:local_effect-semi-infinite-line},  proves the left inequality.
%

\section{Proof of Lemma~\ref{lem:R_min=1/2}}
\label{app:R_min=1/2}

When~$0<pt \ll 1$, a Taylor expansion of~\eqref{eq:f_1D=lim} yields
$\label{eq:S_1D_explicit_expansion}
f^{\rm 1D}(t;p,q) \sim 1-e^{-pt- \frac{qpt^2}{2}}$.
If, in addition, $0<qpt^2 \ll 1$, then
\begin{equation}
	\label{eq:S_1D_explicit_expansion2}
	f^{\rm 1D}(t;p,q) \sim pt+ \frac{qpt^2}{2}.
\end{equation}

Next, let $\frac{q}{p} \gg 1$ and let $t_{\alpha}:=p^{-\frac{1}{1+\alpha}}q^{-\frac{\alpha}{1+\alpha}}$ for some~$\alpha>1$. Then~$qp t_{\alpha}^2 = p^{\frac{\alpha-1}{\alpha+1}}q^{\frac{1-\alpha}{\alpha+1}} =  (\frac{q}{p})^{-\frac{\alpha-1}{\alpha+1}} \ll 1$, $pt_{\alpha} = p^{\frac{\alpha}{1+\alpha}}q^{-\frac{\alpha}{1+\alpha}} = (\frac{q}{p})^{-\frac{\alpha}{1+\alpha}}\ll 1$, and~$qt_{\alpha}  = p^{-\frac{1}{1+\alpha}}q^{\frac{1}{1+\alpha}}=\left(\frac{q}{p}\right)^{\frac{1}{1+\alpha}} \gg 1$.
Hence, by~Lemma~\ref{lem:local_effect-semi-infinite-line} and~\eqref{eq:S_1D_explicit_expansion2},
$$
R_{\min} \le R(t_{\alpha})
\sim \frac{pt_{\alpha}+\frac{qpt^2_{\alpha}}{4}}{pt_{\alpha}+\frac{qpt^2_{\alpha}}{2}}
=
\frac{1+\frac{qt_{\alpha}}{4}}{1+\frac{qt_{\alpha}}{2}}
\sim
\frac{\frac{qt_{\alpha}}{4}}{\frac{qt_{\alpha}}{2}} = \frac12.
$$
Since~$R_{\min}\geq \frac12$, see Lemma~\ref{lem:R-boundary-1D}, the result follows.

\section{Dimensional argument}
\label{app:R_min(tilde(q))}

By~\eqref{eq:f_1D=lim}, we can rewrite $f^{\rm 1D}$ using dimensionless variables as $f^{\rm 1D}(t;p,q) = f^{\rm 1D}(\tilde{t};\tilde{q})$, where $\tilde{t}:=pt$
and $\tilde{q}:=\frac{q}p$. Therefore, by~\eqref{eq:R_min-def},
$ R = R(\tilde{t};\tilde{q})$, and so
$
R_{\min}(p,q):=\min_{0 \le t<\infty} R(t;p,q) = \min_{0 \le\tilde{t}<\infty} R(\tilde{t};\tilde{q}) = R_{\min}(\tilde{q}).
$

\section{Proof of Lemma~\ref{lem:f_j-monotone-two-sided-semi-infinite}}
\label{app:f_j-monotone-two-sided-semi-infinite}

Let
network~$\widetilde{\cal N}$ be obtained from the original line by deleting  the edge $\circled{1} \text{-----} \circled{2}$. Then
$
f_j^{\mathbb{Z}^+}
> \widetilde{f_j^{\mathbb{Z}^+}}
= f_{j-1}^{\mathbb{Z}^+}$ for $j \ge 2$.
%
where the inequality follows from the {\em dominance principle} \citep{Bass-boundary-18}.

\section{One-sided line $\stackrel{\longrightarrow}{\mathbb{Z}^+}$}

In the derivation of the decay rate of boundary effects (Appendix~\ref{app:1D_decay_1D}), we begin with the discrete Bass model on the homogeneous one-sided semi-infinite line~$\stackrel{\longrightarrow}{\mathbb{Z}^+}$,
which
is given by
\begin{subequations}
	\label{eqs:one_sided_line_all-isotropic}
	\begin{equation}
		X_j(0)=0, \qquad j\in \mathbb{Z}^+,
	\end{equation}
	and for $j \in \mathbb{Z}^+$, as $\Delta t\to0$,
	\begin{equation}
		\label{eq:1D_one_sided_line-isotropic}
		\mathbb{P} (X_j(t+\Delta t)=1 \, \vert \,  {\bf X}(t))=
		\begin{cases}
			1,&  {\rm if}\ X_j(t)=1,\\
p \, \Delta t, & {\rm if}\ j=1 \ {\rm and} \  X_1(t)=0,
\\
\left(p+q X_{j-1}(t)\right) \Delta t, &  {\rm if} \ j>1 \ {\rm and} \ X_j(t)=0.
		\end{cases}
	\end{equation}
\end{subequations}
The adoption probability $f^{\stackrel{\longrightarrow}{\mathbb{Z}^+}}_{j}:=\mathbb{P} (X_j(t)=1)$
  of node~$j$ on the one-sided line, where~$X_j$ is the solution of~\eqref{eqs:one_sided_line_all-isotropic}, is equal to that of any node on a circle with~$j$ nodes~\citep{Bass-boundary-18}, i.e.,
\begin{equation}
	\label{eq:Lemma4.4}
	f^{\stackrel{\longrightarrow}{\mathbb{Z}^+}}_{j}(t;p,q)= f^{\rm circle}(t;p,q,j).
\end{equation}

\subsection{Artificial boundary condition}
\label{sec:ABC}

	Consider the discrete Bass model on the {\em $p_1$-heterogeneous}
one-sided line~$\stackrel{\longrightarrow}{\mathbb{Z}^+}$, where

\begin{subequations}
	\label{eqs:one_sided_line-p1-heterogeneous}
	\begin{equation}
		X_j(0)=0, \qquad j\in {\mathbb{Z}^+},
	\end{equation}
	and for $j \in {\mathbb{Z}^+}$, as $\Delta t\to0$,
	\begin{equation}
		\label{eq:one_sided_line-p1-heterogeneous}
		\mathbb{P} (X_j(t+\Delta t)=1 \, \vert \,  {\bf X}(t))=
		\begin{cases}
			1,&  {\rm if}\ X_j(t)=1,
			\\
			p_1(t) \, \Delta t, & {\rm if}\ j=1 \ {\rm and} \  X_1(t)=0,
			\\
\left(p+q X_{j-1}(t)\right) \Delta t, &  {\rm if} \ j>1 \ {\rm and} \ X_j(t)=0.
		\end{cases}
	\end{equation}
\end{subequations}
We can set $p_1$ so that the adoption probabilities of all nodes
on~$\stackrel{\longrightarrow}{\mathbb{Z}^+}$ would be identical to those on the {\em homogeneous} infinite line:

%

\begin{lemma}
	\label{lem:ABC_1s}
	Let~${f}^{\stackrel{\longrightarrow}{\mathbb{Z}^+}}_{j}$ denote the adoption  probability of node~$j$ in the discrete Bass model~\eqref{eqs:one_sided_line-p1-heterogeneous}. If
	\begin{equation}
	{p}_1 = p+q\left(1-e^{-pt}\right),
\end{equation}
then ${f}^{\stackrel{\longrightarrow}{\mathbb{Z}^+}}_{j}(t) \equiv f^{\rm 1D}(t;p,q)$ for $j \in \mathbb{Z}^+$,
	where~$f^{\rm 1D}$ is given by~\eqref{eq:f_1D=lim}.
\end{lemma}
\proof{Proof.}
Let $[S_j]:=1- {f}^{\stackrel{\longrightarrow}{\mathbb{Z}^+}}_{j}$ and $[S^{\rm 1D}] :=1-f^{\rm 1D}$.
We prove that $[S_j] \equiv [S^{\rm 1D}]$ by induction on~$j$.
	Recall that $[S^{\rm 1D}]$ is the solution of \citep{OR-10}
\begin{equation}
	\label{eq:S_1D_der_ABC}
	\frac{d}{dt}{[S^{\rm 1D}]}=-\left(p+q\left(1-e^{-pt}\right)\right)[S^{\rm 1D}], \qquad [S^{\rm 1D}](0) = 1.
\end{equation}
	On the $p_1$-heterogeneous line~\eqref{eq:one_sided_line-p1-heterogeneous},
the master equation for~$[S_1]$ is, see~\citep{fibich2022exact}.
$$
\frac{d}{dt}[S_1] = -p_1 [S_1], \qquad [S_1](0)=1,
$$
 Since  $p_1= p+q(1-e^{-pt})$,
then $[S_1] \equiv [S^{\rm 1D}]$, see~\eqref{eq:S_1D_der_ABC}.

 Assume by induction that $[S_{j-1}] \equiv [S^{\rm 1D}]$.
The master equations for the interior nodes on the $p_1$-heterogeneous line are~\citep{fibich2022exact},
	\begin{equation*}
		\frac{d}{dt}[S_{j}](t) + \left(p+q\right)[S_{j}]  = qe^{-pt}[S_{j-1}],
		\qquad [S_j](0) = 1, \qquad j   = 2,3, \dots
	\end{equation*}
%
%
%
%
%
Substituting $[S_{j-1}] \equiv [S^{\rm 1D}]$ shows that
$[S_j]$ satisfies~\eqref{eq:S_1D_der_ABC}. Hence,  $[S_j] \equiv [S^{\rm 1D}]$.
 \halmos

\section{Super-exponential decay of boundary effects ($1D$)}
\label{app:1D_decay_1D}

We begin with some auxiliary lemmas.

\begin{lemma}
\label{lem:lower_bound_for_any_net}
For any $p,q,t>0$ and any network, $[S](t;p,q) <  e^{-pt}$.
\end{lemma}
\proof{Proof.}
When~$q=0$, all nodes are isolated.
 The non-adoption probability of an isolated node is~$[S] = e^{-pt}$,
see e.g.,~\cite[eq.~(3.8)]{Bass-boundary-18}.
Hence, by the {\em dominance principle}~\citep{Bass-boundary-18},
$
[S](t;p,q) < [S](t;p,q=0) =  e^{-pt}
$.
 \halmos

\begin{lemma}
	\label{lem:sequence_T}
	Let~$\{t_m\}_{m=1}^J$ be $J$ independent identically distributed random variables, such that
	$t_m \sim {\rm exp} (q)$. Let $t>0$.  Then
	\begin{equation}
		\label{eq:lower_upper_bounds}
		\mathbb{P} \left(\sum_{m=1}^J t_m < t\right) < e^{-q t} \left(\frac{eq t}{J} \right)^J, \qquad J>qt.
	\end{equation}
\end{lemma}
\proof{Proof.}
	Let~$\epsilon > 0$. Then
$
		\mathbb{E}\left[e^{-\epsilon t_m}\right] = \int_{0}^{\infty}e^{-\epsilon \tau }\ q e^{-q \tau}{ d\tau}=\frac{q}{\epsilon+q}.
$
	Therefore, by the independence of~$\left\{t_m\right\}_{m=1}^J$, hence of~$\left\{e^{-\epsilon t_m}\right\}_{m=1}^J$,
	\begin{equation}
		\label{eq:mgf_exp2}
		\mathbb{E}\left[e^{-\epsilon\sum_{m=1}^J t_m}\right] = \mathbb{E}\left[\prod_{m=1}^Je^{-\epsilon t_m}\right] = \prod_{m=1}^J \mathbb{E}\left[e^{-\epsilon t_m}\right] = \left(\frac{q}{\epsilon +q}\right)^J.
	\end{equation}
	In addition, for any~$x\in \mathbb{R}$,
	\begin{equation*}
		e^{-\epsilon t} \, \mathbbm{1}_{\left\{x> -\epsilon t\right\}}  < e^{x}.
	\end{equation*}
	Therefore, setting~$x = -\epsilon \sum_{m=1}^J t_m$, and noting that~$\mathbbm{1}_{\left\{-\epsilon\sum_{m=1}^J t_m> -\epsilon t\right\}}=\mathbbm{1}_{\left\{\sum_{m=1}^J t_m< t\right\}}$, yields
	\begin{equation*}
		e^{-\epsilon t} \, \mathbbm{1}_{\left\{\sum_{m=1}^J t_m< t\right\}}   < e^{-\epsilon \sum_{m=1}^J t_m}.
	\end{equation*}
	Taking the expectation of both sides and using~\eqref{eq:mgf_exp2} gives~$e^{-\epsilon t} \, \mathbb{P} \left(\sum_{m=1}^J t_m< t\right)   < \left(\frac{q}{\epsilon +q}\right)^J$,
	and so
	\begin{equation}
		\label{eq:upper_bound_eps}
		\mathbb{P} \left(\sum_{m=1}^J t_m< t\right)   < e^{\epsilon t}\left(\frac{q}{\epsilon +q}\right)^J, \qquad \epsilon>0.
	\end{equation}
	
	Inequality~\eqref{eq:upper_bound_eps} was derived for all $\epsilon>0$.
	Substituting $\epsilon = 0$ yields the trivial upper bound~$\mathbb{P} \left(\sum_{m=1}^J t_m< t\right)<1$, and so~\eqref{eq:upper_bound_eps} holds for $\epsilon \geq 0$.
	Hence, 
	\begin{equation*}
		\mathbb{P} \left(\sum_{m=1}^J t_m< t\right)   < \min\limits_{\epsilon \ge  0}
		e^{\epsilon t}\left(\frac{q}{\epsilon +q}\right)^J.
	\end{equation*}
	
	Standard calculus yields
$		\epsilon_{\rm min}:=\arg\min_{\epsilon \geq  0 }e^{\epsilon t} \left(\frac{q}{\epsilon+q}\right)^J = \frac{J}{t}-q.
$
	Since~$0<t\leq \frac{J}{q}$, then $\epsilon_{\rm min} \geq 0$, as needed.
	Plugging~$\epsilon_{\rm min}$ in~\eqref{eq:upper_bound_eps} gives~\eqref{eq:lower_upper_bounds}.
\halmos

\subsection{One-sided line}

We now prove the super-exponential decay of boundary effects on the one-sided line:
\begin{lemma}
	\label{lem:one-sided_decay}
	Consider the discrete Bass
	model~\eqref{eqs:one_sided_line_all-isotropic} on the homogeneous one-sided line~$\stackrel{\longrightarrow}{\mathbb{Z}^+}$.
	Let $p,q>0$.
	Then for any $t>0$,
	\begin{equation}
		\label{eq:one-sided_decay}
		0< f^{\rm 1D}(t;p,q) - f^{\stackrel{\longrightarrow}{\mathbb{Z}^+}}_{j}(t;p,q)<e^{-\left(p+q\right)t}\left(\frac{eqt}{j}\right)^j,
		\qquad   j \ge qt.
	\end{equation}
\end{lemma}
\proof{Proof.}
Let~$f_j:=f^{\stackrel{\longrightarrow}{\mathbb{Z}^+}}_{j}$,
$\overline{f_{j}}:=\overline{f}^{\stackrel{\longrightarrow}{\mathbb{Z}^+}}_{j}$, and~$\widehat{f}_j:=\widehat{f}^{\stackrel{\longrightarrow}{\mathbb{Z}^+}}_{j}$
denote the adoption probability of node~$j$  in~\eqref{eqs:one_sided_line-p1-heterogeneous} with
%
 $p_1 = p$,
%
with $\overline{p}_1 = p+q$,
and with $\widehat{p}_1 =	p+q\left(1-e^{-pt}\right)$,
respectively.
Since $p_1<\widehat{p}_1<\overline{p}_1$ and all other rate parameters are identical,
then
by the {\em dominance principle}~\citep{Bass-boundary-18},
$$
f_{j}(t)
<\widehat{f}_{j}(t)
<\overline{f_{j}}(t),
\qquad j \in \mathbb{Z}^+,
\quad t>0.
$$
In addition, by Lemma~\ref{lem:ABC_1s},
$\widehat{f}_{j}(t) \equiv f^{\rm 1D}(t)$
for $j  \in \mathbb{Z}^+$. Hence,
$$
f_{j}(t)
<f^{\rm 1D}(t)
<\overline{f}_{j}(t),
\qquad j \in \mathbb{Z}^+,
\quad t>0.
$$
Therefore, to prove~\eqref{eq:one-sided_decay}, it is sufficient to show that
\begin{equation}
	\label{eq:one-sided_decay_suff}
	\overline{f_{j}}(t;p,q) - f_{j}(t;p,q)<e^{-\left(p+q\right)t}\left(\frac{eqt}{j}\right)^j, \qquad  j \ge qt.
\end{equation}

To prove~\eqref{eq:one-sided_decay_suff}, we define stochastic realizations of
the discrete Bass model~\eqref{eqs:one_sided_line-p1-heterogeneous} on the $p_1$-heterogeneous line as follows.
Let~$t_k = k\Delta t$.
A specific realization~$\widetilde{X}_m(t_k)$ of~${X}_m(t_k)$ for nodes~$m \in \mathbb{Z}^+$ is defined as follows:
\renewcommand\labelitemii{$\bullet$} 		
\renewcommand\labelitemiii{$\bullet$}
\renewcommand\labelitemiv{$\bullet$}
\begin{itemize}
	\item ${\widetilde{X}}_{m}(0) = 0$ for $m \in \mathbb{Z}^+$
	\item for $k=0,1,\ldots$
	\begin{itemize}
		\item sample a random vector $\protect {\vomega}^{k} = \left(\omega_{1}^{k},\omega_{2}^{k},\ldots\right)$ from the uniform distribution on~$\left[0,1\right]^{\mathbb{Z}^+}$
		\item for $m =  1,2,\ldots$
		 \begin{itemize}
			\item if ${\widetilde{X}}_{m}(t_{k}) = 1$, then ${\widetilde{X}}_{m}(t_{k+1}) = 1$,
			\item if ${\widetilde{X}}_{m}(t_{k}) = 0$, then
			\begin{itemize}
				\item if $m=1$ and  $0 \leq \omega_{1}^{k} \leq  p_1(t_k) \Delta t$, then ${\widetilde{X}}_{1}(t_{k+1}) = 1$, else ${\widetilde{X}}_{1}(t_{k+1}) = 0$
					\item if $m>1$ and $0 \leq \omega_{m}^{k} \leq \left( p + q {\widetilde{X}}_{m-1}(t_{k}) \right)\Delta t$, then ${\widetilde{X}}_{m}(t_{k+1}) = 1$,
			 else ${\widetilde{X}}_{m}(t_{k+1}) = 0$
			\end{itemize}
		\end{itemize}
		\item end
	\end{itemize}
	\item end
\end{itemize} 	

Let us denote by $X_{j}$
and~$\overline{X}_{j}$  the solutions of~\eqref{eqs:one_sided_line-p1-heterogeneous} with $p_1= p$ and with $\overline{p}_1 = p+q$, respectively.
By the {\em dominance principle} \citep{Bass-boundary-18},
$\mathbb{P}\left(\overline{X}_j(t)  \ge   X_j(t) \right) = 1$.
Therefore,
$$
\overline{f_{j}}(t;p,q) - f_{j}(t;p,q)
= \mathbb{P}\left(\overline{X}_j(t) >  X_j(t) \right).
$$		
Hence, to prove~\eqref{eq:one-sided_decay_suff}, we
need to show that
\begin{equation}
	\label{eq:one-sided_decay_suff-X}
	\mathbb{P}\left(\overline{X}_j(t) >  X_j(t) \right) <e^{-\left(p+q\right)t}\left(\frac{eqt}{j}\right)^j, \qquad  j \ge qt.
\end{equation}

To do that, we first note that
since $p_k \equiv \overline{p}_k$ and  $q_k \equiv \overline{q}_k$ for $k \ge 2$,
%
if any node $k_0 \in \mathbb{Z}^+$ adopts at the same time in both networks, then nodes~$\{k_0+1, k_0+2, \dots \}$ will also adopt at the same time in both networks.
Therefore,  a \emph{necessary condition} for the event
$\overline{X}_j(t) >  X_j(t)$
is that
\begin{enumerate}
	\item Node~$j$ in the homogeneous line does not adopt by external influences until time~$t$, and
	\item  there is an increasing sequence of times $0<T_{k_1}^{\rm het}<T_{k_2}^{\rm het}<\cdots < T_{k_j}^{\rm het} <t$, such that
	for $m \in \{1,\ldots, j\}$,  node~$m$ adopts at time~$T_{k_m}^{\rm het} = k_m^{\rm het} \Delta t$ in the heterogeneous line, before it adopts in the homogeneous line.
\end{enumerate}
This necessary condition (this condition is not sufficient, because it does not guarantee that node~$j$ does not adopt by time~$t$ in the homogeneous line.
	For example, assume that in addition to this condition,
(i)~nodes~$\left\{1,\ldots,j-2\right\}$  do not adopt in the homogeneous line  by time~$t$, (ii)~node~$j-1$ adopts in the homogeneous line by external influences
		at time~$T_{k_{j-1}}^{\rm hom}$,  where $ T_{k_{j}}^{\rm het}<T_{k_{j-1}}^{\rm hom}<t$, and (iii)~node~$j$ adopts in the homogeneous line
		by internal influences from~$j-1$ at time~$T_{k_{j}}^{\rm hom}$, where $T_{k_{j-1}}^{\rm hom}<T_{k_{j}}^{\rm hom}<t$.
	Then node~$j$ adopts by time~$t$ in both networks.)
	holds only if
\begin{subequations}
	\label{eqs:p*Delta_t<omega_m^k_m<=(p+q)Delta_t}
	\begin{equation}
		\label{eq:p*Delta_t<omega_m^k_j<=1}
		p\Delta t<\omega_j^{k}\leq 1, \qquad 0 \leq k  \le \frac{t}{\Delta t},
	\end{equation}
	and  there is an increasing sequence of integers $0<k_1<k_2<\cdots < k_j <\frac{t}{\Delta t}$, such that
	\begin{equation}
		\label{eq:p*Delta_t<omega_m^k_m<=(p+q)Delta_t}
		p\Delta t<\omega_m^{k_m} \leq (p+q)\Delta t, \qquad m = 1,\ldots, j.
	\end{equation}
\end{subequations}

Define the random variable~$T^{\rm hom}_j (\cdot):=\inf\{t \, | \,t = k \Delta t, \, 0 \le \omega_j^k \le p \Delta t\}$ to be the earliest time at which
node~$j$ in the homogeneous line adopts by external influences,
and the random variables $t^{\rm het}_m(\cdot):=
T_{k_m}^{\rm het}-T_{k_{m-1}}^{\rm het}$ for $m=1,\ldots,j$, where $t_{k_0}^{\rm het}:=0$.
Then condition~\eqref{eqs:p*Delta_t<omega_m^k_m<=(p+q)Delta_t} can be written as
$\left\{ T^{\rm hom}_j > t, ~\sum_{m=1}^jt^{\rm het}_m < t \right\}$, and so
\begin{equation*}
	\mathbb{P} \left(\text{condition~\eqref{eqs:p*Delta_t<omega_m^k_m<=(p+q)Delta_t}} \right) =
	\mathbb{P} \left(T^{\rm hom}_j > t, ~\sum_{m=1}^jt^{\rm het}_m < t \right ).
\end{equation*}

\begin{itemize}
	\item By~\eqref{eq:p*Delta_t<omega_m^k_j<=1},
	$T^{\rm hom}_j$ is geometrically distributed with parameter
	$
	1- \mathbb{P} \left(p\Delta t<\omega_j^{k} \leq 1\right)
	= p\Delta t.
	$
	Therefore,
	$T^{\rm hom}_j \sim \exp (p)$  as $\Delta t \to 0$, and so
	$\lim_{\Delta t \to 0} \mathbb{P} (T^{\rm hom}_j > t) = e^{-pt}$.
	
	\item By~\eqref{eq:p*Delta_t<omega_m^k_m<=(p+q)Delta_t}, for~$m=1,\ldots,j-1$,
	$t^{\rm het}_m \mid (T^{\rm hom}_j >t)$ is geometrically distributed with parameter
	$$
	\mathbb{P} \left(p\Delta t<\omega_m^{k} \leq \left(p+q\right)\Delta t\right)
	= q\Delta t.
	$$
	Hence, $t^{\rm het}_m \mid (T^{\rm hom}_j >t) \sim \exp (q)$ as $\Delta t \to 0$.
	
	\item By~\eqref{eq:p*Delta_t<omega_m^k_j<=1} and~\eqref{eq:p*Delta_t<omega_m^k_m<=(p+q)Delta_t}, $t^{\rm het}_j \mid (T^{\rm hom}_j >t)$ is geometrically distributed with parameter
	$$
	\mathbb{P} \left(p\Delta t<\omega_j^{k} \leq \left(p+q\right)\Delta t\mid p\Delta t<\omega_j^{k}\leq 1\right)
	= \mathbb{P} \left(0\leq \omega_j^{k} \leq \frac{q\Delta t}{1-p\Delta t} \right)
	= q\Delta t+O\left(\Delta t^2\right).
	$$
	Hence, $t^{\rm het}_j  \mid (T^{\rm hom}_j >t) \sim \exp (q)$
	as $\Delta t \to 0$.
	
\end{itemize}
Combining the above and using the upper bound~\eqref{eq:lower_upper_bounds}
gives
\begin{equation*}
	\begin{aligned}
		\mathbb{P}\left(\overline{X}_j(t) >  X_j(t) \right)
		&\le
		\lim_{\Delta t \to 0}
		\mathbb{P} \left(\text{condition~\eqref{eqs:p*Delta_t<omega_m^k_m<=(p+q)Delta_t}} \right)
		=
		\lim_{\Delta t \to 0}
		\mathbb{P} \left(T^{\rm hom}_j > t, \sum_{m=1}^jt^{\rm het}_m < t \right )
		<
		e^{-pt} 	e^{-qt} \left(\frac{e q t}{j} \right)^j,
	\end{aligned}
\end{equation*}

%
%
%
%
which is~\eqref{eq:one-sided_decay_suff-X}.
\halmos

From Lemma~\ref{lem:one-sided_decay}, we can obtain a spatial estimate
for $f^{\rm 1D} - f^{\stackrel{\longrightarrow}{\mathbb{Z}^+}}_{j}$,
 which is {\em globally uniform in time}, that shows that boundary effects
decay exponentially in~$j$:
\begin{corollary}
	\label{cor:one-sided_decay-spatial-only}
	Assume the conditions of Lemma~\ref{lem:one-sided_decay}. Then
	\begin{equation}
		\label{eq:one-sided_decay-spatial-only}
		0< f^{\rm 1D}(t;p,q) - f^{\stackrel{\longrightarrow}{\mathbb{Z}^+}}_{j}(t;p,q)<	\left(\frac{q}{p+q}\right)^j, \qquad  0< t <\infty, \quad j \in \mathbb{Z}^+.
	\end{equation}	
\end{corollary}
\proof{Proof.}
	The maximum of $e^{-(p+q)t} t^j$ in $0 \le t<\infty$ is attained at $t_m := \frac{j}{p+q}$.
	Since $0<t_m<\frac{j}{q}$,
	\begin{equation}
		\label{eq:one-sided_decay-small-times}
		\sup_{0 \le t \le \frac{j}{q}}  e^{-(p+q)t}\left(\frac{eqt}{j}\right)^j
		=  e^{-(p+q)t_m}\left(\frac{eqt_m}{j}\right)^j =
		e^{-j} \left(\frac{eq}{p+q}\right)^j = \left(\frac{q}{p+q}\right)^j.
	\end{equation}
	To extend this estimate to $0\le  t<\infty$, we recall that by Lemma~\ref{lem:lower_bound_for_any_net},
	$$
	\left| f^{\rm 1D}(t;p,q) - f^{\stackrel{\longrightarrow}{\mathbb{Z}^+}}_{j}(t;p,q)  \right|  = \left| [S^{\stackrel{\longrightarrow}{\mathbb{Z}^+}}_{j}](t;p,q) -[S^{\rm 1D}](t;p,q)  \right|
	< e^{-pt}, \qquad 0<t<\infty.
	$$
	Hence,
	\begin{equation}
		\label{eq:one-sided_decay-large-times}
		\sup_{\frac{j}{q} \le  t <\infty} \left|f^{\rm 1D}(t;p,q) - f^{\stackrel{\longrightarrow}{\mathbb{Z}^+}}_{j}(t;p,q) \right| < e^{-p\frac{j}{q}}
		=\left(e^{-\frac{p}{q}}\right)^j.
	\end{equation}
	Let $x >0$. Then $e^x>1+x$ and so $e^{-x}<\frac{1}{1+x}$.
	Substituting $x = \frac{p}{q}$ gives $e^{-\frac{p}{q}}<\frac{q}{p+q}$.
	Therefore, by~\eqref{eq:one-sided_decay-large-times},
	\begin{equation*}
		\sup_{\frac{j}{q} \le  t <\infty} \left|f^{\rm 1D}(t;p,q) - f^{\stackrel{\longrightarrow}{\mathbb{Z}^+}}_{j}(t;p,q) \right| <\left(\frac{q}{p+q}\right)^j.
	\end{equation*}
	The result follows from this inequality and~\eqref{eq:one-sided_decay-small-times}.
\halmos

\subsection{Exponential convergence of~$f^{\rm circle}$ to~$f^{\rm 1D}$ }

Using Lemma~\ref{lem:one-sided_decay}, we can show that the rate of convergence
of~$f^{\rm circle}$ to~$f^{\rm 1D}$ is exponential in~$M$:
\begin{lemma}
	\label{lem:circle_diff}
    Let $p,q>0$. Then for any $t>0$,
	\begin{equation}
		\label{eq:f_circle_diff}
		0<f^{\rm 1D}(t;p,q)-f^{\rm circle}(t;p,q,M)< e^{-(p+q)t}\left(\frac{eqt}{M}\right)^M, \qquad  M > qt.
	\end{equation}
	In addition,
\begin{equation}
	\label{eq:exponentail-convergence-f_circle-boundary-spatial}
	0<f^{\rm 1D}(t;p,q) - f^{\rm circle}(t;p,q,M) <
	\left(\frac{q}{p+q}\right)^M, \qquad 0 < t <\infty, \qquad M=1,2, \dots
\end{equation}
\end{lemma}
\proof{Proof.} This result follows from relations~\eqref{eq:f_1D=lim},
 \eqref{eq:Lemma4.4}, and~\eqref{eq:one-sided_decay}. \halmos

\subsection{Two-sided line}

We can now prove the  rate of decay of boundary effects on the two-sided line:

\proof{\bf Proof of Theorem~\ref{thm:1D_decay}}
	By~\eqref{eq:S_j_semi_infinite} and~\eqref{eq:S_1D=S_1D^2},  
\begin{equation}
	\label{eq:proof_line}
	\begin{aligned}
		f^{\rm 1D}(t;p,q) - f^{\mathbb{Z}^+}_{j}(t;p,q)
		& =  [S^{\mathbb{Z}^+}_{j}](t;p,q) -[S^{\rm 1D}](t;p,q)
		\\
		&= e^{p t}[S^{\rm circle}]\left(t;p,\frac{q}{2},j \right)
		[S^{\rm 1D}]\left(t;p,\frac{q}{2} \right) - e^{pt}[S^{\rm 1D}]^2 \left(t;p,\frac{q}{2} \right)
		\\
		&=e^{pt}[S^{\rm 1D}]\left(t;p,\frac{q}{2} \right)\left(f^{\rm 1D}\left(t;p,\frac{q}{2} \right)-f^{\rm circle}\left(t;p,\frac{q}{2},j \right)\right).
	\end{aligned}
\end{equation}
Since~$[S]<e^{-pt}$,  
see Lemma~\ref{lem:lower_bound_for_any_net}, then
\begin{equation}
	\label{eq:s_monotone_finite}
	e^{pt}[S^{\rm 1D}]\left(t;p,\frac{q}{2}\right) < 1.
\end{equation}
Therefore, the result follows from~\eqref{eq:f_circle_diff}, \eqref{eq:proof_line}, 
and~\eqref{eq:s_monotone_finite}.
\halmos



\proof{\bf Proof of~Corollary~\ref{cor:two-sided_decay-spatial-only}}
This follows from~\eqref{eq:Lemma4.4}, \eqref{eq:one-sided_decay-spatial-only},  \eqref{eq:proof_line}, and~\eqref{eq:s_monotone_finite}.

\proof{\bf Proof of Lemma~\ref{lem:1D_decay_finite}}
Combining relations~\eqref{eq:S_j_funnel} and~\eqref{eq:S_1D=S_1D^2} gives
\begin{equation}
\label{eq:s_finite-s_1D}
\begin{aligned}
[S_j^{[1,\ldots,M]}]&(t;p,q)-[S^{\rm 1D}](t;p,q) = e^{pt}\left([S^{\rm circle}]\left(t;p,\frac{q}{2},j \right)
[S^{\rm circle}]\left(t;p,\frac{q}{2},M+1-j \right) - [S^{\rm 1D}]^2\left(t;p,\frac{q}{2} \right) \right)
\\
& = e^{pt}[S^{\rm 1D}]\left(t;p,\frac{q}{2}\right)\left([S^{\rm circle}]\left(t;p,\frac{q}{2},j\right)-[S^{\rm 1D}]\left(t;p,\frac{q}{2}\right)\right)\\
&+e^{pt}[S^{\rm circle}]\left(t;p,\frac{q}{2},j\right)\left([S^{\rm circle}]\left(t;p,\frac{q}{2},M+1-j\right)-[S^{\rm 1D}]\left(t;p,\frac{q}{2}\right)\right).
\end{aligned}
\end{equation}
Therefore, \eqref{eq:1D_decay_finite} follows from~\eqref{eq:f_1D=lim}, \eqref{eq:s_monotone_finite}, \eqref{eq:s_finite-s_1D}, Lemma~\ref{lem:circle_diff}, Lemma~\ref{lem:lower_bound_for_any_net}, and~$[S]=1-f$.
The proof of~\eqref{eq:1D_decay_finite-global-in-t} is similar to that of
Corollary~\ref{cor:two-sided_decay-spatial-only}.  \halmos


For future reference, we derive a space-time estimate for $f^{\rm 1D}- f^{\stackrel{\longrightarrow}{\mathbb{Z}^+}}_{j}$, which is valid for all times and for all~$j$, and decays
exponentially in time and in space:
\begin{lemma}
	Assume the conditions of Lemma~\ref{lem:one-sided_decay}. Then
	\begin{equation}
		\label{eq:one-sided_decay-spatial-only-exer}
		0< f^{\rm 1D}(t;p,q) - f^{\stackrel{\longrightarrow}{\mathbb{Z}^+}}_{j}(t;p,q)<
		e^{-\frac{p}{2} t}	\left(\frac{q}{\frac{p}{2}+q}\right)^j, \qquad  0< t <\infty, \quad j=1,2, \dots
	\end{equation}	
\end{lemma}
\proof{Proof.}

By~\eqref{eq:one-sided_decay},
\begin{equation*}
	0< 	e^{\frac{p}{2} t} \left(f^{\rm 1D}(t;p,q) - f^{\stackrel{\longrightarrow}{\mathbb{Z}^+}}_{j}(t;p,q) \right)  <
	e^{-\left(\frac{p}{2}+q\right)t}\left(\frac{eqt}{j}\right)^j, \qquad  0< t \le \frac{j}{q}.
\end{equation*}
By~\eqref{eq:one-sided_decay-small-times},
\begin{equation*}
	\sup_{0 \le t \le \frac{j}{q}}  e^{-(\frac{p}{2}+q)t}\left(\frac{eqt}{j}\right)^j
	= \left(\frac{q}{\frac{p}{2}+q}\right)^j.
\end{equation*}
by Lemma~\ref{lem:lower_bound_for_any_net},
$$
e^{\frac{p}{2} t} \left(\left| f^{\rm 1D}(t;p,q) - f^{\stackrel{\longrightarrow}{\mathbb{Z}^+}}_{j}(t;p,q) \right) \right| < e^{-\frac{p}{2}t}, \qquad 0<t<\infty.
$$
Hence,
\begin{equation*}
	\sup_{\frac{j}{q} \le  t <\infty} e^{\frac{p}{2} t}\left|f^{\rm 1D}(t;p,q) - f^{\stackrel{\longrightarrow}{\mathbb{Z}^+}}_{j}(t;p,q) \right|
	< e^{-\frac{p}{2}\frac{j}{q}}
	=\left(e^{-\frac{\frac{p}{2}}{q}}\right)^j.
\end{equation*}
Since $e^{-\frac{\frac{p}{2}}{q}}<\frac{q}{\frac{p}{2}+q}$,
\begin{equation*}
	\sup_{\frac{j}{q} \le  t <\infty} e^{\frac{p}{2} t}\left|f^{\rm 1D}(t;p,q) - f^{\stackrel{\longrightarrow}{\mathbb{Z}^+}}_{j}(t;p,q) \right|
	<
	\left(\frac{q}{\frac{p}{2}+q}\right)^j.
\end{equation*}
Hence, the result follows. \halmos

\section{Proof of Theorem~\ref{thm:global_effect2-boundary}}
  \label{app:global_effect2-boundary}

Let $D^M_j:=2f^{\mathbb{Z}^+}_j - f^{[1,\ldots,M]}_j-f^{\rm 1D}$. Then
\begin{equation}
		\label{eq:f^[1,ldots,M]-f^1D}
	f^{[1,\ldots,M]}-f^{\rm 1D}
	= \frac{1}{M}\sum_{j=1}^M\left(f^{[1,\ldots,M]}_j-f^{\rm 1D} \right)
	= \frac{2}{M}\sum_{j=1}^M\left(f^{\mathbb{Z}^+}_{j}-f^{\rm 1D}\right) -  \frac{1}{M}\sum_{j=1}^MD_j^M.
\end{equation}
We claim that
\begin{equation}
	\label{eq:im_sum_D_j^M}
\lim\limits_{M\rightarrow \infty}\sum_{j=1}^MD_j^M=0.
\end{equation}
Therefore, letting $M \to \infty$ in~\eqref{eq:f^[1,ldots,M]-f^1D} and using~\eqref{eq:global_effect_psi-1D} and~\eqref{eq:im_sum_D_j^M}
proves the theorem.

To prove~\eqref{eq:im_sum_D_j^M}, we use relations \eqref{eq:S_j_funnel}, \eqref{eq:S_j_semi_infinite}, \eqref{eq:S_1D=S_1D^2},  and the identities  $[S]:= 1- f$
and
	\begin{equation}
		\label{eq:sum_symm}
		\sum_{j=1}^M[S^{\rm circle}](t;p,q,j) = \sum_{j=1}^M[S^{\rm circle}](t;p,q,M+1-j),
	\end{equation}
		to get
	\begin{equation*}
		\begin{aligned}
			\sum_{j=1}^M D_j^M &=
			\sum_{j=1}^M \bigg[ [S^{[1,\ldots,M]}_j](t;p,q)+[S^{\rm 1D}](t;p,q) -2[S^{\mathbb{Z}^+}_j](t;p,q)\bigg] \\
			&   = e^{pt}\sum_{j=1}^M\bigg[[S^{\rm circle}]\left(t;p,\frac{q}{2},j\right)[S^{\rm circle}]\left(t;p,\frac{q}{2},M+1-j\right)+[S^{\rm 1D}]^2 \left(t;p,\frac{q}{2}\right)\\
			& \qquad \qquad \qquad -2[S^{\rm circle}]\left(t;p,\frac{q}{2},j\right)[S^{\rm 1D}]\left(t;p,\frac{q}{2}\right)\bigg]\\
			& \overset{\mathrm{\eqref{eq:sum_symm}}}{=} e^{pt}\sum_{j=1}^M\bigg[[S^{\rm circle}]\left(t;p,\frac{q}{2},j\right)[S^{\rm circle}]\left(t;p,\frac{q}{2},M+1-j\right)+[S^{\rm 1D}]^2 \left(t;p,\frac{q}{2}\right)\\
			& \qquad \qquad \qquad -  [S^{\rm circle}]\left(t;p,\frac{q}{2},j\right)[S^{\rm 1D}]\left(t;p,\frac{q}{2}\right)
			-[S^{\rm circle}]\left(t;p,\frac{q}{2},M+1-j\right)[S^{\rm 1D}]\left(t;p,\frac{q}{2}\right)\bigg]\\
			& = e^{pt}\sum_{j=1}^M\left([S^{\rm circle}]\left(t;p,\frac{q}{2},j\right)-[S^{\rm 1D}]\left(t;p,\frac{q}{2}\right)\right)\left([S^{\rm circle}]\left(t;p,\frac{q}{2},M+1-j\right)-[S^{\rm 1D}]\left(t;p,\frac{q}{2}\right)\right).
		\end{aligned}
	\end{equation*}
	In addition, by~\eqref{eq:one-sided_decay-spatial-only-exer},
	$$
	0< e^{\frac{p}{2}t} \left([S^{\rm circle}]\left(t;p,\frac{q}{2},j\right)-[S^{\rm 1D}]\left(t;p,\frac{q}{2}\right) \right) <
	\left(\frac{\frac{q}{2}}{\frac{p}{2}+\frac{q}{2}}\right)^j
	=\left(\frac{q}{p+q}\right)^j.
	$$
	Therefore,
	%
	\begin{equation*}
		\begin{aligned}
			0<	\sum_{j=1}^M D_j^M<
			\sum_{j=1}^M \left(\frac{q}{p+q} \right)^j \left(\frac{q}{p+q} \right)^{M+1-j}
			= M \left(\frac{q}{p+q}\right)^{M+1}.
		\end{aligned}
	\end{equation*}
	Since $0<\frac{q}{p+q}<1$, the result follows.


\section{Proof of Lemma~\ref{lem:psi-1D}}
\label{app:psi-1D}

From~\eqref{eq:exponentail-convergence-f_circle-boundary-spatial}
and~\eqref{eq:proof_line}   it follows that
\begin{equation}
	\label{eq:f_1D-f_j<s^pt*S_1D}
	0< f^{\rm 1D}(t;p,q) - f^{\mathbb{Z}^+}_{j}(t;p,q)< e^{pt}[S^{\rm 1D}]\left(t;p,\frac{q}{2}\right) \left(\frac{\frac{q}{2}}{p+\frac{q}{2}}\right)^j, \qquad  0< t <\infty.
\end{equation}	
Hence
	\begin{equation}
			\label{eq:global_effect_psi-bounds}
		0<\underbrace{\sum_{j=1}^{\infty}\left(f^{\rm 1D}(t;p,q)-f^{\mathbb{Z}^+}_{j}(t;p,q)\right)}_{=\psi}
		<
		e^{pt}[S^{\rm 1D}]\left(t;p,\frac{q}{2}\right) \sum_{j=1}^{\infty}\left(\frac{\frac{q}{2}}{p+\frac{q}{2}}\right)^j =
		e^{pt}[S^{\rm 1D}]\left(t;p,\frac{q}{2}\right) \frac{q}{2p}.
	\end{equation}
	\begin{enumerate}
		\item Since $e^{pt}[S^{\rm 1D}]\left(t;p,\frac{q}{2}\right)<1$, see Lemma~\ref{lem:lower_bound_for_any_net}, the first item follows from~\eqref{eq:global_effect_psi-bounds}.

		\item Substituting $f^{\rm 1D}(t=0) = f_j^{\mathbb{Z}^+}(t=0) = 1$
		in~\eqref{eq:global_effect_psi-bounds} gives $\psi(t=0)=0$. In addition,
	      by~\eqref{eq:f_1D=lim},	
			\begin{equation}
			\label{eq:e^pt*S_1D-decays-exponentially}
			e^{pt}[S^{\rm 1D}]\left(t;p,\frac{q}{2}\right) = e^{-\frac{q}{2}\left(t -\frac{1-e^{-pt}}{p}\right)} < e^{-\frac{q}{2}\left(t-\frac1p\right)}, \qquad t > 0.
		\end{equation}	
	    Therefore, $\lim_{t \to \infty} \psi = 0$  by~\eqref{eq:global_effect_psi-bounds} and~\eqref{eq:e^pt*S_1D-decays-exponentially}.

		\item This follows from the first item. 	
		
	\end{enumerate}

\section{Proof of Lemma~\ref{lem:f^N*Z_j1_j2}}
\label{app:f^N*Z_j1_j2}

Independence in~${\bf j}_{-1}$ follows from translation invariance.
To prove monotonicity in~$j_1$, note that
if network~$\widetilde{\cal N}$ is obtained from the original network by deleting the (influential) edges between $(1,{\bf j}_{-1})$ and $(2,{\bf j}_{-1})$ for
all  ${\bf j}_{-1} \in \mathbb{Z}^{D-1}$, then
$
f^{\mathbb{Z}^+ \times \mathbb{Z}^{D-1}}_{(j_1,{\bf j}_{-1})}
> \widetilde{{f}^{\mathbb{Z}^+ \times \mathbb{Z}^{D-1}}_{(j_1,{\bf j}_{-1})}}
= f^{\mathbb{Z}^+ \times \mathbb{Z}^{D-1}}_{(j_1-1,{\bf j}_{-1})}$
for $j_1 \ge 2$ and ${\bf j}_{-1} \in \mathbb{Z}^{D-1}$,
where the inequality follows from the {\em dominance principle} \citep{Bass-boundary-18}.

	\section{Proof of Lemmas~\ref{lem:Dtwo-sided} and~\ref{lem:finite_D_decay}}
\label{app:2Dtwo-sided}



We first prove an auxiliary combinatorial identity:

\begin{lemma}
	\label{lem:C_M^L}
	Let~${C}_M^L$ denote the number of paths in~$\mathbb{Z}^{D}$
	with~$L$~nodes  (and $L-1$~edges) that start from the hyperplane
	\begin{equation}
		\label{eq:hyperplane-B_j_-1}
		H_{{\bf j}_{1} \equiv 1}:=\left\{(1,{\bf j}_{-1}) \, | \, {\bf j}_{-1} \in \mathbb{Z}^{D-1}\right\},
	\end{equation}
	end at~$(M,{\bf 0})$,  where
	${\bf 0}:=\underbrace{(0, \dots, 0)}_{\times D-1}$, and can have repeated vertices and edges.
	Then
	\begin{subequations}
		\label{eq:C_M^L_bound_num_path}
		\begin{equation}
			{C}_M^L \le \frac{(2(D-1) L)^{k}}{k!}, \qquad 1 \le M \le L<\infty,
		\end{equation}
		where
		\begin{equation}
			k:=L-M.
		\end{equation}
	\end{subequations}
\end{lemma}
\proof{Proof}
We prove~\eqref{eq:C_M^L_bound_num_path} by induction on~$k$. Let
$k=0$.  Since $L=M \ge 1$, the only
$M$-node path is
$(1,{\bf 0}) \to(2,{\bf 0})\to \dots \to (M,{\bf 0}),$  and so
$ C_{M}^{M}=1.
$
Since $\frac{(2(D-1) L)^{k}}{k!}\Big|_{k=0} = 1$,
\eqref{eq:C_M^L_bound_num_path}~holds.

To proceed, note that any path  with $L+1$ nodes from~$H_{{\bf j}_{1} \equiv 1}$ to~$(M,{\bf 0})$
is a path with $L$~nodes from~$H_{{\bf j}_{1} \equiv 1}$ to one of the $D$~neighbors of~$(M,{\bf 0})$, and a final edge from that neighbor to~$(M,{\bf 0})$.
The $D$ neighbors of~$(M,{\bf 0})$ are $(M -1,{\bf 0})$, $(M+1,{\bf 0})$, and the $2(D-1)$ nodes $\{(M, \pm {\bf e}_i) \,| \, i=2, \dots, D\}$.
Hence,
\begin{equation}
	\label{eq:C_M^L+1-induction}
	C_{M}^{L+1} =	C_{M-1}^{L}  +	C_{M+1}^{L}+ 2(D-1) C_{M}^{L}.
\end{equation}


 Let
$k=1$. Then  $C_{M}^{L}=C_{M}^{M+1}$.
Substituting $L = M$ in~\eqref{eq:C_M^L+1-induction}, we have
$$
C_{M}^{M+1} =	C_{M-1}^{M}  +\underbrace{C_{M+1}^{M}}_{=0} + 2(D-1) C_{M}^{M} = C_{M-1}^{M}+ 2(D-1).
$$
Therefore, by reverse induction in~$M$,
$$
C_{M}^{M+1} = C_{M-1}^{M}+ 2(D-1) = C_{M-2}^{M-1}+ 4(D-1) = \dots =\underbrace{C_{1}^{2}}_{=2(D-1)}+ (M-1)(D-1)= 2M(D-1).
$$
Thus,
$
C_{M}^{M+1} = 2M(D-1).
$
Since the right-hand-side of~\eqref{eq:C_M^L_bound_num_path} for $L = M+1$ is
$\frac{(2(D-1) L)^{k}}{k!}\Big|_{k=1} = 2(D-1)(M+1)$,
inequality~\eqref{eq:C_M^L_bound_num_path} is satisfied for  $L = M+1$.

For the induction stage, assume that~\eqref{eq:C_M^L_bound_num_path} holds for
$k \in \{ 0,1, \dots, k_0\}$, i.e., for  $1 \le M \le L \le M+k_0$.
We prove that~\eqref{eq:C_M^L_bound_num_path}  holds for  $1 \le M \le L \le M+k_0+1$, as follows.
Substituting $L = L_0  = M+k_0$ in~\eqref{eq:C_M^L+1-induction} and using the induction assumption,
\begin{equation*}
	\begin{aligned}
		C_{M}^{L_0+1}
		& \le
		\frac{(2(D-1) L_0)^{k_0+1}}{(k_0+1)!}
		+ \frac{(2(D-1)L_0)^{k_0-1}}{(k_0-1)!}+  2(D-1) \frac{(2(D-1) L_0)^{k_0}}{(k_0)!}
		\\ & = \frac{(2(D-1) L_0)^{k_0+1}}{(k_0+1)!}
		\left(1 + \frac{ k_0(k_0+1)}{(2(D-1))^2}\frac1{L_0^2} + \frac{k_0+1}{L_0}\right).
	\end{aligned}
\end{equation*}	
Therefore, we need to prove that for $k_0 = 1,2, \dots $,
$$
L_0^{k_0+1}
\left(1 + \frac{k_0(k_0+1)}{(2(D-1))^2}\frac1{L_0^2}  +\frac{k_0+1}{L_0}\right) \le (L_0+1)^{k_0+1},
$$
or
$$
L_0^{k_0+1}+ \frac{ k_0(k_0+1)}{(2(D-1))^2}L_0^{k_0-1}  + (k_0+1) L_0^{k_0}
\le L_0^{k_0+1}+ {k_0+1 \choose 1} L_0^{k_0}+ {k_0+1 \choose 2} L_0^{k_0-1} + \text{positive terms} .	
$$
Hence, the result follows.
\halmos

\begin{remark}
	A proof similar to that of~\eqref{eq:C_M^L_bound_num_path}
	shows that when $D=2$, 
	$	C_{M}^L = {2L-2 \choose L-M}$ for $1\leq M \leq L$.
	\end{remark}

\proof{\bf Proof of Lemma~\ref{lem:Dtwo-sided}}

Any node ${\bf j} \in \mathbb{Z}^+ \times \mathbb{Z}^{D-1}$
can be transformed into a node ${\bf j} \in \mathbb{Z}^D$ by adding edges.
Since these edges are influential to ${\bf j}$,
by the dominance principle \citep{Bass-boundary-18},
$$
0<f^{\rm D}-f^{\mathbb{Z}^+ \times \mathbb{Z}^{D-1}}_{{\bf j}}.
$$

Let~$\widetilde{C}_M^L$ denote the number of paths with~$L\geq M\geq 1$ nodes on~$\mathbb{Z}^+ \times \mathbb{Z}^{D-1}$ that start from the hyperplane~$H_{{\bf j}_1\equiv 1}$, see~\eqref{eq:hyperplane-B_j_-1},
end at~$(M,{\bf 0})$, and do not have repeated vertices. Denote these paths by~$\left\{\gamma^L_n\right\}_{n=1}^{\widetilde{C}_M^L}$,
where $\gamma^L_n =  \circled{\text{${\bf j}^1$}} \longrightarrow\circled{\text{${\bf j}^2$}} \longrightarrow \dots \longrightarrow \circled{\text{${\bf j}^L$}}$.
Let~$\langle \gamma^L_n\rangle $ denote the event that ${\bf j}^k$ adopts in~$\mathbb{Z}^D$ strictly before it adopts in~$\mathbb{Z}^+ \times \mathbb{Z}^{D-1}$ for $k = 1, \dots, L$, and that~$(M, {\bf 0})$  does not adopt from external influence by time~$t$. As in the proof of Lemma~\ref{lem:one-sided_decay}, this event is a necessary
(but not a sufficient)  condition for the event that $(M, {\bf 0})$ adopts by time~$t$
in~$\mathbb{Z}^D$ but not in~$\mathbb{Z}^+ \times \mathbb{Z}^{D-1}$. Therefore, by the union bound,
\begin{equation}
\label{eq:fD_suf_cond-boundary}
f^{\rm D}(t;p,q) - f^{\mathbb{Z}^+ \times \mathbb{Z}^{D-1}}_{(M, {\bf 0})}(t;p,q) \leq \mathbb{P} \left(\cup_{L=M}^{\infty}\cup_{n=1}^{\widetilde{C}_M^L}\langle \gamma^L_n\rangle \right)\leq \sum_{L=M}^{\infty}\sum_{n=1}^{\widetilde{C}_M^L} \mathbb{P} \left(\langle \gamma^L_n\rangle \right).
\end{equation}
We stress that the second inequality in~\eqref{eq:fD_suf_cond-boundary} is a {\em huge overestimate}, since it completely ignores
the dependence between different paths that share some nodes.

Let $t>0$, and let $M \ge \frac{q}{2D}t$. Then $L \ge M \ge  \frac{q}{2D}t$. Since
the internal influence of each adopter on his neighbor is~$\frac{q}{2D}$, then by Theorem~\ref{thm:1D_decay},
\begin{equation}
\label{eq:D_effect_on_path}
\mathbb{P} \left(\left[\gamma^L_n\right]\right) < e^{-\left(p+\frac{q}{2D}\right) t} \left(\frac{e \frac{q}{2D} t}{L} \right)^L. 
\end{equation}
Furthermore, since the paths~$\left\{\gamma^L_n\right\}_{n=1}^{\widetilde{C}_M^L}$
cannot go through the left half-plane $ (-\infty, \dots, 0]\times \mathbb{Z}^{D-1}$, and cannot have repeated vertices,
\begin{equation}
\label{eq:D_bound_num_path}
\widetilde{C}_M^L<C_M^L.
\end{equation}

%

Hence, combining~\eqref{eq:fD_suf_cond-boundary}, \eqref{eq:D_effect_on_path}, and~\eqref{eq:D_bound_num_path} gives 
\begin{equation*}
	\begin{aligned}
		f^{\rm D}(t;p,q) & - f^{\mathbb{Z}^+ \times \mathbb{Z}^{D-1}}_{(M,{\bf 0})}(t;p,q)
		<
		e^{-\left(p+\frac{q}{2D}\right) t} \sum_{L=M}^{\infty} {C}_M^L \left(\frac{e \frac{q}{2D}t}{L} \right)^L
		\\
		& \overset{\eqref{eq:C_M^L_bound_num_path}}{\le}
		e^{-\left(p+\frac{q}{2D}\right) t} \sum_{L=M}^{\infty} 	
		\frac{(2(D-1))^{L-M} L^{L-M}}{(L-M)!} \left(\frac{e \frac{q}{2D}t}{L} \right)^L
		\\
		& <
		e^{-\left(p+\frac{q}{2D}\right) t} \left(\frac{e \frac{q}{2D}t}{M} \right)^M \sum_{L=M}^{\infty} 	
		\frac{(2(D-1))^{L-M} L^{L-M}}{(L-M)!} \left(\frac{e \frac{q}{2D}t}{L} \right)^{L-M}
		\\
		& \overset{k:=L-M}{=}
		e^{-\left(p+\frac{q}{2D}\right) t} \left(\frac{e \frac{q}{2D}t}{M} \right)^M \sum_{k=0}^{\infty} 	\frac{(2(D-1))^{k} L^k}{k!}
		\left(\frac{e \frac{q}{2D}t}{L} \right)^k
		\\ &
		e^{-\left(p+\frac{q}{2D}\right) t} \left(\frac{e \frac{q}{2D}t}{M} \right)^M \sum_{k=0}^{\infty} 	\frac{(e \frac{D-1}{D} q t)^k}{k!}
		= e^{-\left(p+\frac{q}{2D}\right) t}  e^{e\frac{D-1}{D}qt} \left(\frac{e \frac{q}{2D}t}{M} \right)^M.
	\end{aligned}
\end{equation*}
\halmos

\proof{\bf Proof of Lemma~\ref{lem:finite_D_decay}}
Any node ${\bf j} \in B_D$
can be transformed into a node ${\bf j} \in \mathbb{Z}^D$ by adding edges.
These edges are influential to ${\bf j}$, and so
$
f^{B_D}_{{\bf j}}<f^{\rm D}
$
by the  dominance principle
\citep{Bass-boundary-18}.

Let~$\partial B_D$ denote the  boundary nodes of~$B_D$.
The proof of the upper bound is similar to that of Lemma~\ref{lem:Dtwo-sided}, where instead of summing the contributions of all the paths from the hyperplane~$H_{{\bf j}_{1} \equiv 1}$ to node~$(M, {\bf 0})$, we sum
the contributions of all the paths from the boundary $\partial B_D$ to~$\bf j$. Using the union bound, see~\eqref{eq:fD_suf_cond-boundary}, and distinguishing between paths that start from different hyperplanes, gives
\begin{equation*}
	\begin{aligned}
		f^{\rm D}-f^{B_D}_{{\bf j}}
		&\leq
		\sum_{i=1}^D	 \left(
		\sum_{\text{paths from}\choose H_{{\bf j}_{i}\equiv 1} \cap B_D} \!\!\!\!\!\! \mathbb{P} \left(\text{adoption path}\right)
		+ \sum_{\text{paths from}\choose H_{{\bf j}_{i}\equiv M_i} \cap B_D} \!\!\!\!\!\! \mathbb{P} \left(\text{adoption path}\right)
		\right).
		%
	\end{aligned}
\end{equation*}
Since the set of paths to~${\bf j}$ from any of the above finite boundaries is a proper subset of the paths to~${\bf j}$  from the corresponding infinite hyperplane, the result follows from Lemma~\ref{lem:Dtwo-sided}.
\halmos

If $\bf j$ lies on the boundary of the
cube~$B_D :=[1,\ldots,M_1]^D$, then as $M \to \infty$,
$\bf j$~will only experiences boundary effects from nodes on the infinite hyperplane on which it lies,
since the effects of nodes on the $2D-1$  other boundaries will vanish in the limit.
Hence, we have the following result:
\begin{lemma}
	\begin{equation}
		\label{eq:lim_M_1->infty-B^M_D}
		\lim_{M\rightarrow \infty} \max_{{\bf j} \in \partial B_D}
		f^{B_D}_{{\bf j}}
		=f^{\mathbb{Z}^+ \times \mathbb{Z}^{D-1}}_{\rm bdry}.
	\end{equation}
\end{lemma}
\proof{Proof}
By rotational symmetry,
$
\max_{{\bf j} \in \partial B_D} f^{B_D}_{{\bf j}}
= \max_{{\bf j} \in H_{{\bf j}_{i}\equiv 1} \cap \partial B_D}  f^{B_D}_{{\bf j}}.
$
Let ${\bf j} \in H_{{\bf j}_{i}\equiv 1} \cap \partial B_D$.
Similarly to the proof of Lemma~\ref{lem:finite_D_decay},
\begin{align*}
	0 & < f^{\mathbb{Z}^+ \times \mathbb{Z}^{D-1}}_{\rm bdry}-
	f^{B_D}_{{\bf j}}
\le 	\sum_{i=2}^{D}	
	\sum_{\text{paths to ${\bf j}$ from}\choose H_{{\bf j}_{i}\equiv 1} \cap B_D} \!\!\!\!\!\! \mathbb{P} \left(\text{adoption path}\right)
	+  \sum_{i=1}^D \sum_{\text{paths  to ${\bf j}$ from}\choose H_{{\bf j}_{i}\equiv M_i} \cap B_D} \!\!\!\!\!\! \mathbb{P} \left(\text{adoption path}\right)
	.
\end{align*}
Hence,
\begin{align*}
	& 0< f^{\mathbb{Z}^+ \times \mathbb{Z}^{D-1}}_{\rm bdry}-
	\max_{{\bf j} \in H_{{\bf j}_{i}\equiv 1} \cap \partial B_D}  f^{B_D}_{{\bf j}}
	\\ &  \le
	\max_{{\bf j} \in H_{{\bf j}_{i}\equiv 1} \cap \partial B_D}  \left(
	\sum_{i=2}^{D}	
	\sum_{\text{paths to ${\bf j}$ from}\choose H_{{\bf j}_{i}\equiv 1} \cap B_D} \!\!\!\!\!\! \mathbb{P} \left(\text{adoption path}\right)
	+  \sum_{i=1}^D \sum_{\text{paths  to ${\bf j}$ from}\choose H_{{\bf j}_{i}\equiv M_i} \cap B_D} \!\!\!\!\!\! \mathbb{P} \left(\text{adoption path}\right) \right)
	.
\end{align*}
By Lemma~\ref{lem:Dtwo-sided}, the limit as $M_1 \to \infty$ of each of the $2D-1$ sums on the right hand-side is zero, uniformly in ${\bf j} \in H_{{\bf j}_{i}\equiv 1} \cap \partial B_D$.	
Therefore, the result follows.
\halmos

\section{Proof of Theorem~\ref{thm:global_effect2-boundary-2D}}
\label{app:global_effect2-boundary-2D}
We prove~\eqref{eq:f_D_global} by showing that there exist functions~$\psi^{\rm min}_{\rm D}$ and~$\psi^{\rm max}_{\rm D}$ such that
\begin{equation*}
	0<\psi^{\rm min}_{\rm D}(t;p,q)< \lim_{M_1\rightarrow \infty} \left(M_1 \left(f^{\rm D}(t;p,q)-f^{B_D}(t;p,q)\right)\right) < \psi^{\rm max}_{\rm D}(t;p,q)<\infty.
\end{equation*}
We begin with~$\psi^{\rm min}_{\rm D}$. We note that
\begin{equation}
	\label{eq:phi_min1}
	\begin{aligned}
		M_1\left(f^{\rm D}-f^{B_D}\right)
		= M_1\left(\frac{1}{M_1^D}\sum_{{\bf j} \in B_D} \left(f^{\rm D}-f^{B_D}_{\bf j}\right)\right)
		=\frac{1}{M_1^{D-1}}\sum_{{\bf j} \in B_D} \left(f^{\rm D}-f^{B_D}_{\bf j}\right).
	\end{aligned}
\end{equation}
Each node~$\bf j$ in the finite D-dimensional cube~$B_D$ can be obtained from the infinite network by edges removal. Hence, by the dominance principle for nodes,
\begin{equation}
	\label{eq:phi_min2}
	f^{\rm D}-f^{B_D}_{\bf j}>0, \qquad {\bf j} \in B_D.
\end{equation}
Combining~\eqref{eq:phi_min1} and \eqref{eq:phi_min2}, we have
\begin{equation*}
	M_1\left(f^{\rm D}-f^{B_D}\right)
	>
	\frac{1}{M_1^{D-1}}\sum_{{\bf j} \in \partial B_D}\left(f^{\rm D}-f^{B_D}_{\bf j}\right) \geq
	\underbrace{\left(\frac{|\partial B_D|}{M_1^{D-1}}\right)}_{\approx 2D}
	\left(f^{\rm D} -\max_{{\bf j} \in \partial B_D} f^{B_D}_{\bf j}\right),
\end{equation*}
where $\partial B_D$ is the boundary of~$B_D$.
Letting $M_1 \rightarrow \infty$ and using~\eqref{eq:lim_M_1->infty-B^M_D} gives
\begin{equation*}
	\lim_{M_1 \rightarrow \infty} M_1\left(f^{\rm D}-f^{B_D}\right)
	\geq 2D \left(f^{\rm D}-f^{\mathbb{Z}^+ \times \mathbb{Z}^{D-1}}_{\rm bdry}\right):=\psi^{\rm min}_{\rm D}>0,
\end{equation*}
where the last inequality follows from Lemma~\ref{lem:f_D_central=f_D} and Corollary~\ref{cor:f_2D_bdry<f_2D_central}.

For~$\psi^{\rm max}_{\rm D}$, let~$U_{\rm D}^M :=\left\{{\bf j}~|~\frac{q}{2D}t \leq j_1, \dots, j_D \leq M_1-\frac{q}{2D}t\right\}$ and $V_{\rm D}^M:=B_{\rm D}^M \setminus U_{\rm D}^M$ denote the inner and boundary-layer nodes, respectively. On the inner nodes, boundary effects
decay exponentially with distance from the boundary (Lemma~\ref{lem:finite_D_decay}).
On the~$ 2D M_1^{D-1} \times \frac{q}{2D}t = M_1^{D-1}qt$ boundary nodes, we can
use the  bound $|f^{\rm D}-f^{B_D}_{{\bf j}}| \le 1$.
Hence, by~\eqref{eq:finite_D_decay} and~\eqref{eq:phi_min1},
\begin{equation*}
	\begin{aligned}
		&M_1\left(f^{\rm D}-f^{B_D}\right)
		= \frac{1}{M_1^{D-1}}\left[\sum_{{\bf j} \in  V_{\rm D}^M}\left(f^{\rm D}-f^{B_D}_{\bf j}\right)+\sum_{{\bf j} \in  U_{\rm D}^M}\left(f^{\rm D}-f^{B_D}_{\bf j}\right)\right]
		\\ & <
		\frac{1}{M_1^{D-1}}\left[M_1^{D-1}qt+	e^{-\left(p+\frac{q}{2D}\right) t}  e^{e\frac{D-1}{D}qt}  \left(\sum_{j_1=1}^{M_1} \cdots \sum_{j_D=1}^{M_1} \right) \sum_{i=1}^D \left(\left(\frac{e\frac{q}{2D}t}{j_i}\right)^{j_i}+\left(\frac{e\frac{q}{2D}t}{M_1-j_i}\right)^{M_1-j_i}\right) \right]
		\\	& = qt+	2D e^{-\left(p+\frac{q}{2D}\right) t}  e^{e\frac{D-1}{D}qt}
		\sum_{j=1}^{M_1} \left(\frac{e\frac{q}{2D}t}{j}\right)^{j}.
	\end{aligned}
\end{equation*}
Since~$\sum\limits_{j=1}^\infty \left(\frac{e\frac{q}{2D}t}{j}\right)^{j}<\infty$,
\begin{equation*}
	\lim_{M_1 \rightarrow \infty} M_1 \left(f^{\rm D}-f^{B_D}\right)\leq qt +2De^{-\left(p+\frac{1-2e}{4}q\right)t}\sum_{j=1}^\infty\left(\frac{e\frac{q}{2D}t}{j}\right)^{j}:=\psi^{\rm max}_{\rm D}<\infty.
\end{equation*}

\end{APPENDICES}

\ACKNOWLEDGMENT{This material is based upon work supported by the U.S. Department of Energy's Office of Energy Efficiency and Renewable Energy (EERE) under the Solar Energy Technologies Office Award Number~DE-EE0007657. The views expressed herein do not necessarily represent the views of the U.S. Department of Energy or the United States Government.}

\ECSwitch


\ECHead{Additional Proofs and Further Details}

\section{Explicit lower bound for $f^{\mathbb{Z}^+ \times \mathbb{Z}^{D-1}}_{\rm bdry}$}
\label{app:f^D_bdry-lower-bound}

   Previously, we introduced the discrete Bass model~\eqref{eqs:one_sided_line_all-isotropic} on a one-sided line.
  We can similarly define {\em one-sided diffusion in $D$~dimensions},
   where peers effect can only be exerted in e.g., the negative direction of each coordinate. The discrete Bass model on the one-sided $D$-dimensional Cartesian network is given by
\begin{subequations}
	\label{eq:Bass-model-D-dimensional-1-sided}
	\begin{equation}
		\label{eq:general_initial-$D$-dimensional-1-sided}
		X_{\bf j}(0)=0, \qquad {\bf j} \in {\Bbb Z}^d,
	\end{equation}
	and for any ${\bf j} \in {\Bbb Z}^d$, as $ \Delta t \to 0$,
	\begin{equation}
		\mathbb{P} (X_{\bf j}(t+\Delta t )=1 \, \vert \,  {\bf X}(t))=
		\begin{cases}
			\left(p+\frac{q}{D}N_{\bf j}^{\rm 1-sided}(t) \right) \Delta t , & {\rm if}\ X_{\bf j}(t)=0,
			\\
			\hfill 1,\hfill & {\rm if}\ X_{\bf j}(t)=1,
		\end{cases}
	\end{equation}
	where
	\begin{equation}
		\label{eq:N_j(t)-multi-D-1-sided}
		N_{\bf j}^{\rm 1-sided}(t):= \sum_{i=1}^D X_{{\bf j} + \hat{\bf e}_i}(t)
	\end{equation}
	is the number of adopters connected to~${\bf j}$ in the one-sided case.
	%
\end{subequations}

In 1D 	we saw that $f^{\mathbb{Z}^+}_{\rm bdry}(t;p,q)=f^{\rm 1D}\left(t;p,\frac{q}{2}\right)$,
see~\eqref{eq:local_effect-semi-infinite-line}. There is no similar explicit expression for~$f^{\mathbb{Z}^+ \times \mathbb{Z}^{D-1}}_{\rm bdry}$. In particular, we have
\begin{lemma}
	Let $f^{\rm D, \, 1-sided}(t;p,q)$ denote
	the adoption level in
	the discrete Bass model~\eqref{eq:Bass-model-D-dimensional-1-sided} on the one-sided $D$-dimensional Cartesian grid. Then
\begin{equation}
	\label{eq:f_bdry>f_D(t;p,q/2}
	f^{\mathbb{Z}^+ \times \mathbb{Z}^{D-1}}_{\rm bdry}>f^{\rm D, \, 1-sided}\left(t;p,\frac{q}{2} \right).
\end{equation}
\end{lemma}
\proof{Proof.}
Start from the  discrete Bass model~\eqref{eq:Bass-model-2D+} on~$\mathbb{Z}^+ \times \mathbb{Z}^{D-1}$.
If we delete all the directional edges $ \circled{\bf j} \longrightarrow
\circled{\text{${\bf j} + \hat{\bf e}_i$}}$, where $i = 1, \dots, D$,  we obtain the one-sided $D$-dimensional  discrete Bass model~\eqref{eq:Bass-model-D-dimensional-1-sided}, with~$q$
replaced by~$\frac{q}{2}$.
Since the deleted edges are influential to the boundary nodes $j_1 \equiv 1$, then by the {\em dominance principle}~\citep{Bass-boundary-18},
$$
f^{\mathbb{Z}^+ \times \mathbb{Z}^{D-1}}_{\rm bdry}(t;p,q)
>
{f^{\mathbb{Z}^+ \times \mathbb{Z}^{D-1}, \rm \, 1-sided}_{\rm bdry}}\left(t;p, \frac{q}{2} \right).
$$
If we now extend the one-sided network from~$\mathbb{Z}^+ \times \mathbb{Z}^{D-1}$ to~$\mathbb{Z}^D$, the added edges are noninfluential
to the nodes $j_1 \equiv 1$. Therefore,
$$
{f^{\mathbb{Z}^+ \times \mathbb{Z}^{D-1}, \rm \, 1-sided}_{\rm bdry}}\left(t;p,\frac{q}{2} \right)
=
{f^{\mathbb{Z}^D, \rm \, 1-sided}_{(j_1 = 1,{\bf j}_{-1})}}\left(t;p,\frac{q}{2} \right)
=f^{\rm D, \, 1-sided}\left(t;p,\frac{q}{2} \right),
$$
where the last equality follows from translation invariance.
Hence, the result follows.
\halmos

In order to express inequality~\eqref{eq:f_bdry>f_D(t;p,q/2} using the
two-sided Bass model on $\mathbb{Z}^D$, let us recall
\begin{conjecture}[\cite{Bass-boundary-18}]
	\label{conj:f_D_1-sided=f_D_two-sided}
	The expected adoption level in the discrete Bass models~\eqref{eq:Bass-model-D-dimensional-1-sided} and~\eqref{eqs:Bass-general_D} 	on homogeneous $D$-dimensional
	one-sided and two-sided Cartesian networks on $\mathbb{Z}^D$, respectively, are identical, i.e.,
$f^{\rm D, \, 1-sided}(t;p,q) = f^{\rm D}(t;p,q)$.
\end{conjecture}

Therefore, we have
\begin{corollary}
	Let Conjecture~\ref{conj:f_D_1-sided=f_D_two-sided} hold.
	Then
		$f^{\mathbb{Z}^+ \times \mathbb{Z}^{D-1}}_{\rm bdry}>f^{\rm D}\left(t;p,\frac{q}{2} \right)$.
	\end{corollary}

\section{Global effect of an internal boundary}

\begin{figure}[ht!]
	\begin{center}
		\scalebox{0.5}{\includegraphics{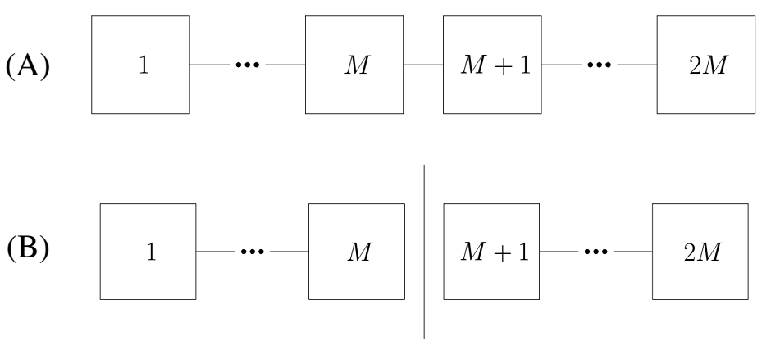}}
		\caption{ Line of length~$2M$ without~(A) and with~(B) an internal boundary.}
		\label{fig:1D-middle_boundary_M_2M}
	\end{center}
\end{figure}

The global effect of internal boundaries is also $O(\frac1M)$.
To see that, we compute the difference~$h_{M}$ between the expected adoption level~$f^{[1,\ldots,2M]}$ on a two-sided line of length~$2M$ (Figure~\ref{fig:1D-middle_boundary_M_2M}A), and the expected adoption level $f^{[1,\ldots,M \, | \, M+1,\ldots,2M]}$ on a two-sided line of length~$2M$ that has  an internal boundary between nodes~$M$ and~$M+1$ (Figure~\ref{fig:1D-middle_boundary_M_2M}B).
Obviously, $f^{[1,\ldots,M \, | \, M+1,\ldots,2M]} \equiv f^{[1,\ldots,M]}$.

\begin{lemma}
	\label{lem:h_M-1D}
	Let
	\begin{equation}
		\label{eq:h_M}
		h_{M}(t;p,q) := f^{[1,\ldots,2M]}(t;p,q) -f^{[1,\ldots,M \, | \, M+1,\ldots,2M]}(t;p,q)
	\end{equation}
	denote the effect of an internal boundary on the aggregate adoption on a two-sided line of length~$2M$.  Then
	\label{lem:global_effect}
	\begin{equation}
		\label{eq:global_effect}
		h_{M}(t;p,q) \sim \frac{\psi(t;p,q)}{M},  \qquad M\rightarrow \infty,
	\end{equation}
	where $\psi$ is given by~\eqref{lem:psi-1D}.
\end{lemma}
\proof{Proof.}
	Since
	$
	f^{[1,\ldots,2M]}-f^{[1,\ldots,M]} =  \left(f^{\rm 1D}-f^{[1,\ldots,M]}\right) -\left(f^{\rm 1D}-f^{[1,\ldots,2M]}\right),
	$
	the result follows from Lemma~\ref{thm:global_effect2-boundary}.
\halmos

\begin{figure}[ht!]	
	\begin{center}
		\scalebox{0.8}{\includegraphics{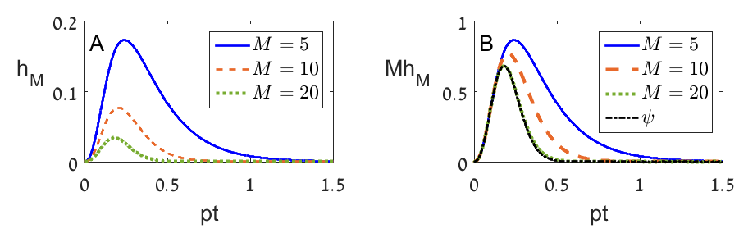}}
		\caption{(A)~Effect of an internal boundary on the adoption level, see~\eqref{eq:h_M}, as a function of time, for various values of~$M$.  Here~$\frac{q}{p}=100$. (B)~Same
			as~(A) with~$Mh_{M}$ as a function of time. The line for~$M=20$ is already indistinguishable from~$\psi$, see~\eqref{lem:psi-1D}.
			  }
		\label{fig:barrier_ratio_onesided_1D_main}
	\end{center}			
\end{figure}

Figure~\ref{fig:barrier_ratio_onesided_1D_main}A
confirms numerically that the effect of an internal boundary decreases with~$M$,
and Figure~\ref{fig:barrier_ratio_onesided_1D_main}B
confirms that $Mh_M \rightarrow \psi$ as~$M \to \infty$, where~$\psi$ is independent of~$M$,
see~\eqref{eq:global_effect}.
%

\section{ Further details on the empirical analysis}
\label{app:empirical}

The primary raw data for the empirical analysis consists of administrative data on residential rooftop solar installations acquired under a non-disclosure agreement from the Connecticut Green Bank. These include the date of the installation and address of the installation, along with several other variables we do not use in this study. The data include nearly all solar installations in Connecticut through the end of~2019. These data are supplemented with U.S.\ Census Bureau data at the Census block group level (a unit that contains between 600 and 3,000 people) from the 2015-2018 American Community Survey. For our study we use median household income, median number of rooms in owner-occupied households, median home values, population, and census block area.

We geocoded each of the addresses in the residential solar installation data using the Google Maps geocoding API (see \url{https://developers.google.com/maps/documentation/geocoding/overview}). This gives a latitude and longitude value for each solar installation. These are then input into ArcMap along with the date of the installation. We next identify each of the municipalities that had a Solarize campaign. Table~\ref{tab:timeline} shows the start and end dates of the~51 municipality-level campaigns included in this analysis. See~\citep{gillingham2021social} for a treatment effects analysis of the effectiveness of the campaigns at spurring further adoption of solar. As mentioned in the main text, we focus on examining the diffusion process after these campaigns because they provide a substantial exogenous increase in solar installations in one area, so we can follow how word-of-mouth and peer effects influence diffusion elsewhere. By focusing on municipalities with a reasonable number of solar installations, we can also improve our signal-to-noise ratio.

We use a Python script run in ArcMap to create buffer zones around each of the Solarize municipality boundaries. These are created within the Solarize municipality itself because we are analyzing the effect of external boundaries. We drop any boundaries where the adjacent municipality is also a Solarize municipality. Then we map the geocoded solar installations to the buffer zones. We next calculate the sum of the installations in each buffer zone in the year following the Solarize campaigns.

For the Census data, we perform a join in ArcMap to overlay the U.S.\ Census block groups geographic areas with the buffers we created. Some block groups may not be entirely within the buffer zone. Thus, we calculate the percentage of the buffer zone covered by each block group and take a weighted average for each Census variable. For example, if there are two block group geographic areas in a buffer zone and 60\% of the block group is covered by one block group and 40\% another, we would take the weighted average based on these percentages. We create the dependent variable for our analysis by dividing the sum of the installations in each buffer zone by the number of owner-occupied households that we calculate are in each buffer zone.


The data for this analysis are at the municipality-buffer zone type level, where a buffer zone type refers to either being in the boundary buffer zone or the inner core zone. Table~\ref{tab:sumstatsinternal} shows the summary statistics for the cross-sectional data with 152 observations, where an observation is a municipality~$i$ zone category $j \in \{\text{boundary buffer}, \text{inner core}\}$. Thus, the analysis includes a total of 76 municipalities. As mentioned above, we excluded the borders where the municipality on the other side of the border is a Solarize municipality (i.e., we examine the border on the other side of the town from the border with the Solarize municipality).

\begin{table}[htbp] \caption{Detailed Timeline of Solarize Campaigns \label{tab:timeline}}	
\begin{center} \begin{adjustbox}{totalheight=\textheight-2\baselineskip}
\begin{tabular}{l c c}
\hline\hline
                        & Start Date    & End Date  \\
\hline
\underline{Round 1}     &               &               \\
Durham                  & Sept 5, 2012  & Jan 14, 2013  \\
Westport                & Aug 22, 2012  & Jan 14, 2013  \\
Portland                & Sept 4, 2012  & Jan 14, 2013  \\
Fairfield               & Aug 28, 2012  & Jan 14, 2013  \\
                        &               &               \\
\underline{Round 2}     &               &               \\
Bridgeport              & Mar 26, 2013  & July 31, 2013 \\
Coventry                & Mar 30, 2013  & July 31, 2013 \\
Canton                  & Mar 19, 2013  & July 31, 2013 \\
Mansfield               & Mar 11, 2013  & July 31, 2013 \\
Windham                 & Mar 11, 2013  & July 31, 2013 \\
                        &               &               \\
\underline{Round 3}     &               &               \\
Easton                  & Sept 22, 2013 & Feb 9, 2014   \\
Redding                 & Sept 22, 2013 & Feb 9, 2014   \\
Trumbull                & Sept 22, 2013 & Feb 9, 2014   \\
Ashford                 & Sept 24, 2013 & Feb 11, 2014  \\
Chaplin                 & Sept 24, 2013 & Feb 11, 2014  \\
Hampton                 & Sept 24, 2013 & Feb 11, 2014  \\
Pomfret                 & Sept 24, 2013 & Feb 11, 2014  \\
Greenwich               & Oct 2, 2013   & Feb 18, 2014  \\
Newtown                 & Sept 24, 2013 & Feb 28, 2014  \\
Manchester              & Oct 3, 2013   & Feb 28, 2014  \\
West Hartford           & Sept 30, 2013 & Feb 18, 2014  \\
West Haven              & Nov 13, 2013  & Apr 8, 2013   \\
Hamden                  & Nov 18, 2013  & Feb 11, 2014  \\
Easton                  & Sept 22, 2013 & Feb 9, 2014   \\
Trumbull                & Sept 22, 2013 & Feb 9, 2014   \\
                        &               &               \\
\underline{Round 4}     &               &               \\
Tolland                 & Apr 23, 2014  & Sept 16, 2014 \\
Torrington              & Apr 24, 2014  & Sept 16, 2014 \\
Simsbury                & Apr 29, 2014  & Sept 23, 2014 \\
Essex                   & Apr 29, 2014  & Sept 23, 2014 \\
Montville               & May 1, 2014   & Sept 23, 2014 \\
Brookfield              & May 6, 2014   & Sept 30, 2014 \\
Bloomfield              & May 6, 2014   & Sept 30, 2014 \\
Farmington              & May 14, 2014  & Oct 7, 2014   \\
Haddam                  & May 15, 2014  & Oct 7, 2014   \\
Killingworth            & May 15, 2014  & Oct 7, 2014   \\
East Lyme               & May 22, 2014  & Oct 14, 2014  \\
Weston                  & June 24, 2014 & Nov 14, 2014  \\
                        &               &               \\
\underline{Round 5}     &               &               \\
Avon                    & Nov 20, 2014  & Apr 10, 2015  \\
Griswold                & Dec 8, 2014   & Apr 28, 2015  \\
Milford                 & Dec 3, 2014   & Apr 23, 2015  \\
Southbury               & Nov 19, 2014  & Apr 9, 2015   \\
Old Lyme                & Dec 4, 2014   & Apr 24, 2015  \\
Lyme                    & Nov 18, 2014  & Apr 8, 2015   \\
South Windsor           & Nov 10, 2014  & Mar 31, 2015  \\
Woodstock               & Dec 3, 2014   & Apr 23, 2015  \\
Burlington              & Nov 19, 2014  & Apr 9, 2015   \\
East Granby             & Dec 2, 2014   & Apr 22, 2015  \\
Suffield                & Dec 2, 2014   & Apr 22, 2015  \\
Windsor                 & Dec 2, 2014   & Apr 22, 2015  \\
Windsor Locks           & Dec 2, 2014   & Apr 22, 2015  \\
New Canaan              & Dec 2, 2014   & Apr 22, 2015  \\
New Hartford            & Nov 17, 2014  & Apr 7, 2015   \\
\hline\hline
\end{tabular}
\end{adjustbox}
\end{center}
\end{table}

\begin{table}[htb]
\protect \caption{\label{tab:sumstatsinternal}Summary statistics for boundary analysis}
\centering \begin{adjustbox}{max width=\textwidth}
\begin{threeparttable}
  \centering
\begin{tabular}{l c c c c }\hline\hline
        & Mean & S.D. & Min. & Max.  \\
\hline
Percent installed (\%)             & 1.17	& 1.04	 &	0.00    & 	7.66  \\
Median household income (\$)       & 98,897 & 31,411 & 56,288   & 241,800 \\
Median number of rooms in homes    & 6.46   & 0.836  & 4.36     &   9.00   \\
Median home values (\$)            & 374,874 & 192,696  & 175,877 & 998,264 \\
Density (owner-occupied houses/area) & 83.7  & 84.0     & 7.55  & 467.4     \\
\hline\hline
\end{tabular}
\begin{tablenotes}[flushleft]
\item \footnotesize \textit{Notes}: There are~152 observations for all of the variables (no missing values), where an observation is a municipality x zone pair. Data are from one year after the period listed for each of the Solarize campaigns in Table~\ref{tab:timeline}. All dollars are nominal dollars. \\
\end{tablenotes}
\end{threeparttable}
\end{adjustbox}
\end{table}

\begin{table}[htbp]
	\protect \caption{\label{tab:internal}Evidence of Boundary Effects}
	\centering \begin{adjustbox}{max width=\textwidth}
		\begin{threeparttable}
			\centering
			\begin{tabular}{l c c c c c c}\hline\hline
				Boundary Buffer                 &  -0.405    &  -0.416   & -0.413   \\
	                                            &  (0.166)   &  (0.156)  &  (0.107)  \\
				Control: Density                &            &     -0.002   &  -0.001   \\
				                                &            &   (0.001)   &  (0.002) \\
				Control: Income                 &            &  -5.52e-06   &  -0.00001   \\
				                                &            &  (4.81e-06) &  (9.81e-06) \\
				Control: Rooms                  &            &  0.483    &  0.865  \\
				                                &            &  (0.141)    & (0.467)  \\
				Control: Home Value             &            &  -1.81e-06  &  9.88e-07   \\
				                                &            &  (6.97e-07) &  5.40e-06 \\
				Constant                        &  1.372    &  -0.326     &  -2.60    \\
	                                            &  (0.141)  &  (0.723)    &  (2.40)  \\
				\hline
				Municipality Fixed Effects      & No         & No        & Yes   \\
				\hline
				R-squared     & 0.038  &  0.185   & 0.825    \\
				N             & 152    & 152      & 152      \\
				\hline\hline
			\end{tabular}
			\begin{tablenotes}[flushleft]
				\item \footnotesize \textit{Notes}: Dependent variable is market share of owner-occupied homes that have installations one year after the Solarize campaigns (mean=0.1). An observation is a municipality x zone pair. Robust standard errors in parentheses. \\
			\end{tablenotes}
		\end{threeparttable}
	\end{adjustbox}
\end{table}

\begin{table}[htbp]
	\protect \caption{\label{tab:empiricalrobust}Robustness Check: Empirical Results Using Non-Solarize Towns}
	\centering \begin{adjustbox}{max width=\textwidth}
		\begin{threeparttable}
			\centering
			\begin{tabular}{l c c c c c c}\hline\hline
				Boundary Buffer                 &  -0.084    &  -0.084   &   -0.084  \\
	                                            &  (0.524)   &  (0.228)  &  (0.040)  \\
				Constant                        &   0.862    &  -15.70     &  23.57    \\
	                                            &  (0.359)  &  (4.42)    &  (2.22)  \\
				\hline
                Control Variables               & No         & Yes       & Yes   \\
				Municipality Fixed Effects      & No         & No        & Yes   \\
				\hline
				R-squared     & 0.002   &  0.874    &   0.997    \\
				N             & 14      & 14        & 14       \\
				\hline\hline
			\end{tabular}
			\begin{tablenotes}[flushleft]
				\item \footnotesize \textit{Notes}: Dependent variable is market share of owner-occupied homes that have installations one year after the Solarize campaigns (mean=0.1). An observation is a municipality x zone pair. The municipalities included are: Bristol, Hamden, Milford, New Haven, Stonington, Stratford, and Waterbury, and the analysis is performed during the period of Round 3, when none of these municipalities had received a Solarize campaign. Robust standard errors in parentheses. \\
			\end{tablenotes}
		\end{threeparttable}
	\end{adjustbox}
\end{table}


\begin{thebibliography}{}

\bibitem[{Albert et~al.(2000)Albert, Jeong, \protect\BIBand{}
  Barab\'asi}]{Albert-00}
Albert R, Jeong H, Barab\'asi A (2000) Error and attack tolerance of complex
  networks. \emph{Nature} 406:378--382.

\bibitem[{Alm \protect\BIBand{} Deijfen(2015)}]{alm2015first}
Alm SE, Deijfen M (2015) First passage percolation on $\mathbb{Z}^2$: A simulation
  study. \emph{Journal of statistical physics} 161:657--678.

\bibitem[{Anderson \protect\BIBand{} May(1992)}]{Anderson-92}
Anderson R, May R (1992) \emph{Infectious Diseases of Humans} (Oxford: Oxford
  University Press).

\bibitem[{Barton-Henry et~al.(2021)Barton-Henry, Wenz, \protect\BIBand{}
  Levermann}]{barton2021decay}
Barton-Henry K, Wenz L, Levermann A (2021) Decay radius of climate decision for
  solar panels in the city of fresno, usa. \emph{Scientific reports} 11:1--9.

\bibitem[{Bass(1969)}]{Bass-69}
Bass F (1969) A new product growth model for consumer durables.
  \emph{Management Sci.} 15:1215--1227.

\bibitem[{Bollinger et~al.(2020)Bollinger, Burkhardt, \protect\BIBand{}
  Gillingham}]{bollinger2020peer}
Bollinger B, Burkhardt J, Gillingham K (2020) Peer effects in residential water
  conservation: Evidence from migration. \emph{American Economic Journal:
  Economic Policy} 12:107--33.

\bibitem[{{Bollinger} \protect\BIBand{} {Gillingham}(2012)}]{Bollinger-12}
{Bollinger} B, {Gillingham} K (2012) Peer effects in the diffusion of solar
  photovoltaic panels. \emph{Marketing Science} 31:900--912.

\bibitem[{Carattini et~al.(2018)Carattini, P{\'e}clat, \protect\BIBand{}
  Baranzini}]{carattini2018social}
Carattini S, P{\'e}clat M, Baranzini A (2018) \emph{Social interactions and the
  adoption of solar PV: evidence from cultural borders} (Grantham Research
  Institute on Climate Change and the Environment).

\bibitem[{Cox \protect\BIBand{} Durrett(1981)}]{cox1981some}
Cox JT, Durrett R (1981) Some limit theorems for percolation processes with
  necessary and sufficient conditions. \emph{The Annals of Probability}
  583--603.

\bibitem[{Evans(1993)}]{evans1993random}
Evans JW (1993) Random and cooperative sequential adsorption. \emph{Reviews of
  Modern Physics} 65:1281.

\bibitem[{Fibich(2016)}]{Bass-SIR-model-16}
Fibich G (2016) Bass-{SIR} model for diffusion of new products in social
  networks. \emph{Phys. Rev. E} 94:032305.

\bibitem[{Fibich(2017)}]{Bass-SIR-analysis-17}
Fibich G (2017) Diffusion of new products with recovering consumers. \emph{SIAM
  J. Appl. Math.} 77:1230-1247.

\bibitem[{Fibich \protect\BIBand{} Gibori(2010)}]{OR-10}
Fibich G, Gibori R (2010) Aggregate diffusion dynamics in agent-based models
  with a spatial structure. \emph{Oper. Res.} 58:1450--1468.

\bibitem[{Fibich \protect\BIBand{} Golan(2023)}]{fibich2021diffusion}
Fibich G, Golan A (2023) Diffusion of new products with heterogeneous
  consumers. \emph{Mathematics of  Operations Research} 48:257-287.

\bibitem[{Fibich \protect\BIBand{} Levin(2023)}]{Bass-networkeffect-20}
Fibich G, Levin T (2023) Funnel theorems for spreading on networks. \emph{ArXiv 2308.13034}.

\bibitem[{Fibich et~al.(2019)Fibich, Levin, \protect\BIBand{}
  Yakir}]{Bass-boundary-18}
Fibich G, Levin T, Yakir O (2019) Boundary effects in the discrete {B}ass
  model. \emph{SIAM J. Appl. Math.} 79:914--937.

\bibitem[{Fibich \protect\BIBand{} Nordmann(2022)}]{fibich2022exact}
Fibich G, Nordmann S (2022) Exact description of {SIR}-{B}ass epidemics on 1{D} lattices.
\emph{Discrete and Continuous Dynamical Systems} 42:505-535.

\bibitem[{Gillingham \protect\BIBand{} Bollinger(2021)}]{gillingham2021social}
Gillingham K, Bollinger B (2021) Social learning and solar photovoltaic
  adoption. \emph{Management Science} 67.

\bibitem[{Givoli(2013)}]{givoli2013numerical}
Givoli D (2013) \emph{Numerical methods for problems in infinite domains}
  (Elsevier).

\bibitem[{Graziano et~al.(2019)Graziano, Fiaschetti, \protect\BIBand{}
  Atkinson-Palombo}]{grazianoetal2019}
Graziano M, Fiaschetti M, Atkinson-Palombo C (2019) Peer effects in the
  adoption of solar energy technologies in the united states: An urban case
  study. \emph{Energy Reserach \& Social Science} 48:75--84.

\bibitem[{Graziano \protect\BIBand{} Gillingham(2015)}]{Graziano-15}
Graziano M, Gillingham K (2015) Spatial patterns of solar photovoltaic system
  adoption: {T}he influence of neighbors and the built environment. \emph{J.
  Econ. Geogr.} 15:815--839.

\bibitem[{Hopp(2004)}]{Hopp-04}
Hopp W (2004) Ten most influential papers of management science's first fifty
  years. \emph{Management Sci.} 50:1763--1893.

\bibitem[{Jackson(2008)}]{Jackson-08}
Jackson M (2008) \emph{Social and Economic Networks} (Princeton University Press).

\bibitem[Kiss et~al.(2017)]{kiss2017mathematics}
I.~Z. Kiss, J.~C. Miller, and P.~L. Simon.
\newblock {\em Mathematics of epidemics on networks}.
\newblock Springer, 2017.

\bibitem[{Kraft-Todd et~al.(2018)Kraft-Todd, Bollinger, Gillingham, Lamp,
  \protect\BIBand{} Rand}]{kraft2018credibility}
Kraft-Todd GT, Bollinger B, Gillingham K, Lamp S, Rand DG (2018)
  Credibility-enhancing displays promote the provision of non-normative public
  goods. \emph{Nature} 563:245.

\bibitem[{Mahajan et~al.(1993)Mahajan, Muller, \protect\BIBand{}
  Bass}]{Mahajan-93}
Mahajan V, Muller E, Bass F (1993) New-product diffusion models. Eliashberg J,
  Lilien G, eds., \emph{Handbooks in Operations Research and Management
  Science}, volume~5, 349--408 (North-Holland, Amsterdam).

\bibitem[{Pastor-Satorras \protect\BIBand{}
  Vespignani(2001)}]{Pastor-Satorras-01}
Pastor-Satorras R, Vespignani A (2001) Epidemic spreading in scale-free
  networks. \emph{Phys. Rev. Lett.} 86:3200--3203.

\bibitem[{Rai et~al.(2016)Rai, Reeves, \protect\BIBand{}
  Margolis}]{rai2016overcoming}
Rai V, Reeves DC, Margolis R (2016) Overcoming barriers and uncertainties in
  the adoption of residential solar pv. \emph{Renew. Energ.} 89:498--505.

\bibitem[{Rai \protect\BIBand{} Robinson(2015)}]{rai2015agent}
Rai V, Robinson SA (2015) Agent-based modeling of energy technology adoption:
  Empirical integration of social, behavioral, economic, and environmental
  factors. \emph{Enviorn. Modell. Softw.} 70:163--177.

\bibitem[{Rogers(2003)}]{Rogers-03}
Rogers E (2003) \emph{Diffusion of Innovations} (New York: Free Press), fifth
  edition.

\bibitem[{Shah(2009)}]{Shah-09}
Shah D (2009) \emph{Gossip Algorithms}
(Now Publishers).

\bibitem[{Strang \protect\BIBand{} Soule(1998)}]{Strang-98}
Strang D, Soule S (1998) Diffusion in organizations and social movements: From
  hybrid corn to poison pills. \emph{Annu. Rev. Sociol.} 24:265--290.

\bibitem[{Tsynkov(1998)}]{tsynkov1998numerical}
Tsynkov SV (1998) Numerical solution of problems on unbounded domains. a
  review. \emph{Applied Numerical Mathematics} 27:465--532.

  \bibitem[{Watts \protect\BIBand{} Strogatz(1998)}]{watts1998collective}
Watts DJ, Strogatz SH (1998) Collective dynamics of ‘small-world’ networks.
\emph{Nature}, 393:440--442.

\bibitem[{Wolf(1987)}]{wolf1987wulff}
Wolf DE (1987) Wulff construction and anisotropic surface properties of
  two-dimensional eden clusters. \emph{Journal of Physics A: Mathematical and
  General} 20:1251.

\bibitem[{Wolske et~al.(2018)Wolske, Todd, Rossol, McCall, \protect\BIBand{}
  Sigrin}]{wolske2018accelerating}
Wolske KS, Todd A, Rossol M, McCall J, Sigrin B (2018) Accelerating demand for
  residential solar photovoltaics: Can simple framing strategies increase
  consumer interest? \emph{Glob. Environ. Chang.} 53:68--77.

\end{thebibliography}

\begin{thebibliography}{}


  \bibitem[{Fibich et~al.(2019)Fibich, Levin, \protect\BIBand{}
  Yakir}]{Bass-boundary-18}
Fibich G, Levin T, Yakir O (2019) Boundary effects in the discrete {B}ass
  model. \emph{SIAM J. Appl. Math.} 79:914--937.

\bibitem[{Gillingham \protect\BIBand{} Bollinger(2021)}]{gillingham2021social}
Gillingham K, Bollinger B (2021) Social learning and solar photovoltaic
  adoption. \emph{Management Science} 67:7091--7112.


\end{thebibliography}
\end{document}